\documentclass[aps,prc,amsfonts,preprintnumbers,superscriptaddress,showpacs,nofootinbib]{revtex4-1}
\usepackage{graphicx}
\usepackage{amsmath}
\usepackage{amssymb}
\usepackage{psfrag}
\usepackage{epsfig}
\usepackage{epsf}
\usepackage{float}
\usepackage{slashed}
\usepackage{longtable}
\usepackage{tabulary}
\usepackage[utf8]{inputenc}

\usepackage[hidelinks]{hyperref}
\usepackage[export]{adjustbox}
\usepackage{hyphenat}
\usepackage{booktabs}
\usepackage{setspace}
\usepackage{physics}
\usepackage{siunitx}
\usepackage{upgreek}
\usepackage{dsfont} 
\usepackage{acronym}
\usepackage[greek,english]{babel}
\allowdisplaybreaks

\usepackage{color}

\renewcommand{\vec}[1]{\boldsymbol{#1}}

\usepackage{natbib}

\newcommand{\trfl}[1]{\operatorname{Tr}\left(#1\right)}

\newcommand{\diag}{\operatorname{diag}}

\newcommand{\identity}{\mathds{1}}

\newcommand{\ci}{\textup{i}}

\newcommand{\tensorsigma}{\textrm{\textgreek{s\noboundary}}}
\newcommand{\levicivita}{\textsf{\textgreek{e}}}
\newcommand{\kronecker}{\text{\textgreek{d}}}

\renewcommand{\paulimatrix}{\textsf{\textgreek{t}}}

\newcommand{\pion}{\pi}

\newcommand{\pionmass}{M_{\pion}}
\newcommand{\barepionmass}{M}

\newcommand{\piondecayconstant}{F_{\pion}}

\newcommand{\nucleon}{N}

\newcommand{\nucleonmass}{m_{\nucleon}}

\newcommand{\nucleoncharge}{Q_{\nucleon}}

\newcommand{\deltapart}{\Delta}

\newcommand{\deltamass}{m_{\deltapart}}
\newcommand{\baredeltamass}{\mathring{m}_{\deltapart}}
\newcommand{\deltafield}{\varPsi}

\newcommand{\pindcoupling}{h_A}
\newcommand{\barehA}{h}
\newcommand{\axialcoupling}{g_A}
\newcommand{\baregA}{g}

\newcommand{\baregOne}{\mathring{g}_1}

\begin{document}

	\title{Pion photoproduction in chiral perturbation theory with explicit treatment of the $\boldmath{\deltapart(1232)}$ resonance}

	\author{N.~R\ij neveen}
	\email[]{Email: nora.rijneveen@rub.de}
	\affiliation{Ruhr-Universit\"at Bochum, Fakult\"at f\"ur Physik und
		Astronomie, Institut f\"ur Theoretische Physik II,  D-44780
		Bochum, Germany}
	
	\author{A.~M.~Gasparyan}
	\email[]{Email: ashot.gasparyan@rub.de}
	\affiliation{Ruhr-Universit\"at Bochum, Fakult\"at f\"ur Physik und
		Astronomie, Institut f\"ur Theoretische Physik II,  D-44780
		Bochum, Germany}
	\affiliation{NRC Kurchatov Institute-ITEP, B.~Cheremushkinskaya 25, 117218 Moscow, Russia}
	
	\author{H.~Krebs}
	\email[]{Email: hermann.krebs@rub.de}
	\affiliation{Ruhr-Universit\"at Bochum, Fakult\"at f\"ur Physik und
		Astronomie, Institut f\"ur Theoretische Physik II,  D-44780
		Bochum, Germany}
	
	\author{E.~Epelbaum}
	\email[]{Email: evgeny.epelbaum@rub.de}
	\affiliation{Ruhr-Universit\"at Bochum, Fakult\"at f\"ur Physik und
		Astronomie, Institut f\"ur Theoretische Physik II,  D-44780
		Bochum, Germany}
	
	\date{\today}

	\begin{abstract}
		We study the reaction of pion photoproduction on the
                nucleon in the framework of chiral perturbation theory
                with explicit $\deltapart(1232)$ degrees of
                freedom. In the covariant approach, we give results up
                to order $\epsilon^3$ in the small scale expansion
                scheme. Furthermore, we provide $\deltapart$-less and
                $\deltapart$-full results obtained in the heavy-baryon
                scheme to analyze the differences to the covariant
                approach. Low energy constants are fitted to multipole
                amplitudes using theoretical truncation errors estimated by a Bayesian approach. We also compare our findings to data of neutral pion production cross sections and polarization asymmetries. The description of the reaction is clearly improved by the explicit treatment of the $\deltapart(1232)$ resonance.
	\end{abstract}

	
	\maketitle

\section{Introduction}
In this work, we study the process of pion photoproduction on the
nucleon, i.e.~$\gamma + N\to \pi +N $, in the framework of chiral
perturbation theory ($\chi$PT) with explicit $\deltapart$ degrees of
freedom. Studying this reaction is motivated by several
reasons. First, it is the photoproduction of the lightest hadron. From
the theoretical point of view, it is one of the simplest processes
involving three different particles, and it thus serves as a test field
for more complex reactions. From the experimental point of view, this
reaction is quite easily accessible close to threshold and there can
be no other hadronic final states due to energy
conservation. Therefore,  a lot of experimental data on pion
photoproduction are available.
As an additional motivation, pion photoproduction contributes as a subprocess to more complicated reactions, e.g.,
in radiative pion photoproduction $\gamma + N \to \gamma + \pi + N$, providing an access to 
the magnetic moment of the $\deltapart$ resonance \cite{Rijneveen:2020qbc},  
or in the interaction of nuclei with electromagnetic 
probes \cite{Pastore:2009is, Kolling:2011mt, Kolling:2012cs, Piarulli:2012bn, Schiavilla:2018udt, Krebs:2019aka}.

Pion photoproduction has been a research topic for many years, with
the first model-independent approach proposed by Kroll and Ruderman in
the 1950s \cite{Kroll:1953vq}. Based on general principles, such as
Lorentz and gauge invariance, they derived a low-energy theorem for
the matrix element of charged pion photoproduction at threshold, which
expressed the production amplitudes in terms of a series in the
parameter $\mu=\pionmass/\nucleonmass$, where $\pionmass$ and
$\nucleonmass$ refer to pion and nucleon masses, respectively. Later
on, these predictions were improved
\cite{DeBaenst:1971hp,Vainshtein:1972ih} by including the so-called
partially conserved axial-vector current hypothesis
\cite{Nambu:1960xd,Bernstein:1960qxa, GellMann:1960np} and current
algebra \cite{Book:Adler1968, Book:Treiman1972,
  Book:deAlfaro1973}. Until the 1980s, there was little doubt about
the validity of the low-energy predictions. Particularly for the
charged production channels, which are dominated by the Kroll-Ruderman
term, the results from the theorem matched the available data
well. However, new data for the neutral production channel at
threshold \cite{Mazzucato:1986dz, Beck:1990da} showed a non-negligible
disagreement with the theoretical predictions for the $s$-wave
electric dipole amplitude $E_{0+}$. A first important success for
$\chi$PT was the study of Bernard, Kaiser, Gasser and Meißner
\cite{Bernard:1991rt, Bernard:1992nc} on pion photoproduction, which
corrected the low-energy theorems by terms arising from pion loop
diagrams. However, these corrections, generated by infrared
singularities of the loop integrals, even worsened the agreement with
data, which is due to the slow convergence of the chiral expansion in
the neutral pion production channel. Therefore, renewed interest in
pion photoproduction was awakened in the following years, and multiple
experimental groups remeasured pion photo- and electroproduction
reactions, see e.g.~\cite{Welch:1992ex,
  vandenBrink:1995uka,Blomqvist:1996tx, Fuchs:1996ja,
  Bergstrom:1996fq, Bergstrom:1997jc, Bernstein:1996vf, Kovash:1997tj,
  Distler:1998ae, Liesenfeld:1999mv, Korkmaz:1999sg, Schmidt:2001vg,
  Merkel:2001qg, PhDBaumann, Weis:2007kf, Merkel:2009zz,
  Merkel:2011cf, Hornidge:2012ca, Hornidge:2013qka, Lindgren:2013eta}.
In parallel, Bernard et al.\ worked out all the theoretical details in the different reaction channels within the framework of heavy baryon chiral perturbation theory (HB$\chi$PT) \cite{Bernard:1992qa, Bernard:1992nc, Bernard:1992ys, Bernard:1993bq, Bernard:1994dt, Bernard:1996bi, Bernard:1994gm, Bernard:1996ti, Bernard:1995cj, Fearing:2000uy}. 
An improvement of the convergence of the chiral expansion has been achieved by extending 
the pure $\chi$PT approach by means of dispersion-relation techniques \cite{Gasparyan:2010xz}. 

The new approaches of the so-called infrared renormalization \cite{Becher:1999he} and the extended-on-mass-shell scheme (EOMS) \cite{Fuchs:2003qc} enabled the treatment of scattering processes in the pion-nucleon sector in a manifestly covariant framework. Consequently, covariant calculations of $\gamma +N \to \pi + N$ were completed up to the leading loop order $\order{{Q}^3}$ in Ref.~\cite{Bernard:2005dj} using the infrared renormalization scheme, followed by an analysis of the full one-loop order $\order{{Q}^4}$ by Hilt et al.\ \cite{Hilt:2013fda,Hilt:2013uf} in the EOMS scheme. 

After it was worked out how to include the $\deltapart$ resonance as an explicit degree of freedom into $\chi$PT 
within the so-called small scale expansion (SSE) scheme
\cite{Hemmert:1996xg, Hemmert:1997ye}, the photoproduction of pions
has also been considered in this extended framework. An explicit
treatment of the $\deltapart$ is of great interest, because the small
gap between the pion production threshold and the $\deltapart$ mass
suggests that effects of the $\deltapart$ may become important close above threshold. For example, the $M_{1+}$ multipole receives dominant effects from the $\deltapart$ resonance for energies close to the $\deltapart$ mass.
In SSE, the $\deltapart$-nucleon mass split $\Delta \equiv \deltamass
- \nucleonmass$ is considered to be of the same order
as the pion mass, i.e.~$\Delta\sim\pionmass\sim\epsilon$.
The first calculation of pion photoproduction with explicit $\deltapart$ degrees of freedom was completed in HB$\chi$PT \cite{Hemmert:1996xg}, focusing on the neutral production channel close to threshold, which showed only moderate effects of the explicit $\deltapart$ treatment. A subsequent HB$\chi$PT study \cite{Cawthorne:2015orf} found a more distinct improvement of the description in the HB framework. In the covariant approach, first studies of neutral pion photoproduction were done in Refs.~\cite{Blin:2014rpa, Blin:2016itn, Navarro:2019iqj, Navarro:2020zqn}, finding a substantial improvement in the description of the data compared to the HB approach with explicit $\deltapart$s. However, in these works a different power counting \cite{Pascalutsa:2002pi} is used, the so-called $\delta$ scheme, in which the $\deltapart$-nucleon mass split $\Delta$ is considered to be of one order lower than the pion mass $\Delta \sim \delta, \pionmass \sim \delta^2$. The motivation for such a counting is given by numerical arguments. Since there is no clear evidence for a faster convergence or better efficiency of one of the two schemes, it is desirable to consider pion photoproduction in the SSE for comparison purposes. Note, however, that the low energy constants (LECs) of the two approaches cannot be compared directly, because their numerical values are scheme-dependent. Comparing the results in two counting schemes can be done only in terms of quality of the data description.

We study pion photoproduction in both heavy baryon and covariant
formalisms of chiral effective field theory. We also analyze the
effects of explicit $\deltapart$ degrees of freedom by taking into
account their leading-order and next-to-leading order contributions in
the covariant formalism of $\chi$PT within the EOMS scheme. To estimate theoretical uncertainties of observables and LECs, we use a Bayesian model \cite{Furnstahl:2015rha, Melendez:2017phj, Epelbaum:2019zqc}. 

First, we study the $\deltapart$-less case up to order ${Q}^3$, which has been considered before \cite{Bernard:2005dj, Hilt:2013fda, Hilt:2013uf}. Recent studies were mainly focused on the covariant approach, so we provide a detailed comparison of the two formalisms in order to analyze the difference in convergence and data description. 
In particular, we compare the obtained LECs in terms of the effects generated by the infrared regular (IR) parts of the integrals. 
The extraction of the heavy baryon LECs is very important for the use in the few-body applications
such as calculation of the nuclear electroweak currents.

Next, we upgrade the covariant calculation to leading-order
$\deltapart$ tree contributions employing the SSE, which has been used
for studies of other reactions such as pion-nucleon scattering and
Compton scattering, see e.g.~Refs.~\cite{Siemens:2016hdi,
  Siemens:2016jwj, Thurmann:2020mog}. We also discuss the differences
to the $\deltapart$-less case in terms of resonance
saturation. Moreover, we provide results for the reaction  $\gamma + N \to \pi +N $ up to leading $\deltapart$-full loop order for the first time. In comparison to the study of Ref.~\cite{Blin:2016itn}, we take into account loop diagrams including up to three $\deltapart$ propagators, which give rise to significant contributions to the amplitude while maintaining 
gauge invariance. Furthermore, we give an estimate for the subleading $\gamma N \Delta$ coupling constant $h_1$. A qualitative comparison to the aforementioned studies in the $\delta$-scheme is given. Furthermore, we compare some of our results to recent high-precision data in the neutral pion production channel \cite{Hornidge:2012ca, Hornidge:2013qka, PCHornidge}. 

Our paper is structured as follows. In section \ref{sec:kinematics}, we introduce the basic formalism, such as kinematics, isospin/spin decomposition of the matrix elements and calculation of multipole amplitudes and observables. In section \ref{sec:Lagrangian}, we list 
all terms of the effective Lagrangian relevant for our calculation and
discuss the employed power counting scheme. Subsequently, we discuss renormalization in section \ref{sec:renormalization}. We present our results in section \ref{sec:results} and conclude with a short summary in section \ref{sec:summary}.
\section{Formalism} 
\label{sec:kinematics}
%
%
In this section, we briefly introduce our notation for kinematics, isospin and spin decomposition as well as the calculation of multipole amplitudes and observables. Pion photoproduction 
\begin{align}
	\gamma(\lambda, k) + N_i(s,p) \to \pi_c(q) + N_j(s^\prime, p^\prime)
\end{align}
is a reaction where a pion is produced by absorption of a photon from a nucleon. Here, $p (p^\prime)$, $s (s^\prime)$ and $i (j)$ are momentum, helicity and isospin index of the incoming (outgoing) nucleon, $k$ and $\lambda$ are momentum and helicity of the photon, and $q$ and $c$ are momentum and isospin index of the pion, respectively.
The kinematics of the reaction is uniquely defined by two Lorentz-invariant Mandelstam variables
\begin{align}
s=(p+k)^2=(p^\prime +q)^2, \quad t=(p-p^\prime)^2=(k-q)^2\,.
\end{align}
The energies of the photon $\omega$ and the pion $E_{\pi}$ in the center-of-mass (CM) frame expressed in terms of the Mandelstam variables read
\begin{align}
\omega=\frac{s-\nucleonmass^2}{2\sqrt{s}}, \quad E_{\pi}=\frac{s+\pionmass^2-\nucleonmass^2}{2\sqrt{s}}.
\label{eq:CMquantities}
\end{align}
In the laboratory frame, the photon energy can be calculated from the total CM energy using
\begin{align}
	\omega_{\text{lab}}=\frac{s-\nucleonmass^2}{2\nucleonmass}.
\end{align}
Furthermore, we define the scattering angle $\theta$ via $\lvert \vec{k} \rvert\,\lvert \vec{q} \rvert \cos(\theta) = \vec{k}\cdot\vec{q}$, such that
\begin{align}
	t=\pionmass^2-2(\omega E_{\pi} - \lvert \vec{k} \rvert\,\lvert \vec{q} \rvert \cos(\theta)).
\end{align}
The pion production threshold lies at the CM energy of $\sqrt{s} = \nucleonmass+\pionmass$.
In the HB formalism, an explicit $1/\nucleonmass$ expansion of the amplitude is performed. Expanding Eq.~\eqref{eq:CMquantities} in terms of $1/\nucleonmass$, we can derive an approximate relation between photon and pion CM energy 
\begin{align}
	\omega = E_\pi - \frac{\pionmass^2}{2\nucleonmass} + \order{\frac{1}{\nucleonmass^2}}. 
\end{align}
Thus, in the HB formalism, we express all kinematic quantities in terms of the pion energy $E_\pi$.
%
%

In the isospin space, the matrix elements of pion photoproduction can be parameterized in terms of three independent structures
\begin{align}
	T_{\gamma  N \to \pi  N}^c= \delta_{c3}\, T^{(+)}_{\gamma N} + \tau_c\, T^{(0)}_{\gamma N} + \ci \varepsilon_{c3a}\, \tau_a\, T^{(-)}_{\gamma N}, 
\label{eq:T_Isospin}
\end{align}
where $\tau^i$ are the Pauli matrices in the isospin space. 
In Eq.~\eqref{eq:T_Isospin}, $T$ can refer to the pion photoproduction amplitude in any representation
including the photoproduction multipoles.

The matrix elements for the four physical pion photoproduction reaction channels 
\begin{align}
\gamma p\to \pi^0 p,\quad \gamma p \to \pi^+ n,\quad \gamma n \to \pi^0 n,\quad \gamma n\to \pi^- p
\label{eq:physicalreactionchannels}
\end{align}
can be obtained from the three isospin structures by using the following relations
\begin{align}
\notag 	T_{\gamma p \to \pi^0 p} & = T^{(+)}_{\gamma N} + T^{(0)}_{\gamma N} , & \quad T_{\gamma p \to \pi^+ n}& = \sqrt{2} (T^{(0)}_{\gamma N} + T^{(-)}_{\gamma N}  ),\\
T_{\gamma n \to \pi^0 n}& = T^{(+)}_{\gamma N} - T^{(0)}_{\gamma N} , & \quad T_{\gamma n \to \pi^- p} &=  \sqrt{2} (T^{(0)}_{\gamma N}- T^{(-)}_{\gamma N}).
\label{eq:isospinphysicalreactionchannels}
\end{align}
Another commonly used decomposition of the pion photoproduction amplitude is the so-called isospin parametrization in terms of the three amplitudes $T_{\gamma N}^{(\frac{3}{2})}, T_{\gamma p}^{(\frac{1}{2})}, T_{\gamma n}^{(\frac{1}{2})}$
where the production amplitudes $T_{\gamma N}^{(I)}$ are related to the $T_{\gamma N}^{(0,\pm)}$ via
\begin{align}
	\notag T_{\gamma N}^{(\frac{3}{2})}& = T_{\gamma p}^{(\frac{3}{2})} =T_{\gamma n}^{(\frac{3}{2})}=  T_{\gamma N}^{(+)} - T_{\gamma N}^{(-)}, \\
	T_{\gamma p}^{(\frac{1}{2})}&= T_{\gamma N}^{(0)}+\frac{1}{3} T_{\gamma N}^{(+)} + \frac{2}{3} T_{\gamma N}^{(-)}, \quad T_{\gamma n}^{(\frac{1}{2})}=T_{\gamma N}^{(0)} - \frac{1}{3} T_{\gamma N}^{(+)} - \frac{2}{3}  T_{\gamma N}^{(-)}.
	\label{eq:isospinbasis}
\end{align}
The isospin parametrization is a natural choice when working in the isospin symmetric case of $\chi$PT, thus we perform the fits in this work using this basis. For comparison with experimental results, the decomposition~\eqref{eq:isospinphysicalreactionchannels} in terms of the physical reaction channels is required.

%
%
The pion photoproduction amplitude $\mathcal{M} = \epsilon^{\mu} \mathcal{M}_{\mu}$, where $\epsilon^{\mu}$ is
the polarization vector of the photon, can be parameterized in terms of
the so-called Ball amplitudes \cite{Ball:1961zza}\footnote{We work
  with Hilt's convention, given in Ref.~\cite{Hilt:2013fda}, which is
  slightly different from Ball's original one.}
\begin{align}
\mathcal{M}^{\mu}=\sum\limits_{i=1}^{8} \bar{u}(p^\prime) B_i V_i^{\mu} u(p).
\label{eq:ballamplitudes}
\end{align}
Here the coefficients $B_i$ are scalar functions of the Mandelstam
variables, and the 
basis structures $V_i^\mu$ comprise all independent
matrices that can be formed using gamma matrices and the polarization vector, they read 
\begin{align}
\notag V_1^{\mu}&=\gamma^{\mu} \gamma_5,& V_2^\mu & =  \gamma_5 P^{\mu}, & V_3^{\mu} &= \gamma_5 q^\mu, & V_4^{\mu} &= \gamma_5 k^{\mu},\\
V_5^{\mu} &= \gamma^{\mu}\slashed{k} \gamma_5,& V_6^{\mu} & = \slashed{k} \gamma_5 P^\mu, & V_7^{\mu} &= \slashed{k}\gamma_5 q^{\mu}, & V_8^{\mu}& =\slashed{k} \gamma_5 k^{\mu},
\label{eq:ballamplitudes2}
\end{align}
where $P=1/2\,(p+p^{\prime})$.  Note that the set of amplitudes \eqref{eq:ballamplitudes}-\eqref{eq:ballamplitudes2} is not minimal. Imposing transversality $k_{\mu} \mathcal{M}^\mu=0$ leads to the following conditions
\begin{align}
B_1+B_6\, k\cdot P+B_7\,k\cdot q+ B_8\, k^2=0, \quad B_2\,k\cdot P+B_3\,k\cdot q+B_4\, k^2+B_5\,k^2=0.
\label{eq:ballampsrelation}
\end{align} 
Thus, current conservation reduces the number of basis structures to
six. Only four structures remain for real pion photoproduction due to the additional constraints $\epsilon\cdot k=0$ and $k^2=0$, which can be chosen in the form of CGLN amplitudes \cite{Chew:1957tf}:
\begin{align}
\notag \mathcal{M}^{\mu} & =\sum\limits_{i=1}^{4} \bar{u}(p^\prime) A_i M_i^{\mu} u(p),\\
\notag M_1^{\mu}& =-\tfrac{\ci}{2} \gamma_5(\gamma^{\mu} \slashed{k}- \slashed{k} \gamma^{\mu}),\\
\notag M_2^{\mu}& =2 \ci \gamma_5 \left( P^{\mu} k\cdot (q-\tfrac{1}{2} k) - (q^{\mu}-\tfrac{1}{2} k^{\mu}) k\cdot P \right),\\
\notag M_3^{\mu}&= -\ci \gamma_5 (\gamma^{\mu} k\cdot q - \slashed{k} q^{\mu}),\\
M_4^{\mu}&= -2 \ci \gamma_5 (\gamma^{\mu} k\cdot P - \slashed{k} P^{\mu}) - 2 \nucleonmass M_1^{\mu}.
\label{eq:minimalbasis}
\end{align}
In the CM frame, one can conveniently introduce another set of amplitudes
in the gauge $\epsilon^0=0$ \cite{Chew:1957tf}:
\begin{equation}
\epsilon_{\mu} \bar{u}(p^{\prime}) \left( \sum_{i=1}^{4} A_i M_i^{\mu}
\right) u(p) = \frac{4 \uppi \sqrt{s}}{\nucleonmass}
\chi^{\dagger}_f\, \mathcal{F}\, \chi_i\,,
\label{eq:cglnamplitudes}
\end{equation}
with
\begin{equation}
\mathcal{F}=\ci\, \boldsymbol{\sigma}\,\cdot \boldsymbol{\epsilon}\, \mathcal{F}_1+\frac{\boldsymbol{\sigma}\cdot \vec{q}\, \boldsymbol{\sigma} \cdot (\vec{k}\cross \boldsymbol{\epsilon})}{\abs{\vec{q}}\, \abs{\vec{k}}}\, \mathcal{F}_2+ \ci \frac{\boldsymbol{\sigma} \cdot \vec{k}\, \vec{q}\cdot \boldsymbol{\epsilon}}{\abs{\vec{q}}\, \abs{\vec{k}}}\, \mathcal{F}_3 + \ci \frac{\boldsymbol{\sigma}\cdot \vec{q}\, \vec{q}\cdot \boldsymbol{\epsilon}}{\abs{\vec{q}}^2}\, \mathcal{F}_4,
\end{equation}
where $\boldsymbol{\sigma}$ are the Pauli matrices in spin space
and $\chi_i$ ($\chi^\dagger_f$) is the Pauli spinor
of the initial (final) nucleon. 

The $\mathcal{F}_i$'s can be expanded in a multipole series \cite{Chew:1957tf,Ball:1961zza}:
\begin{align}
\notag \mathcal{F}_1 & = \sum_{l=0}^{\infty}\left\{ [ l\, M_{l+} +E_{l+} ] P^\prime_{l+1}(x) + [ (l+1) M_{l-} + E_{l-} ] P^\prime_{l-1}(x) \right\} ,\\
\notag \mathcal{F}_2 & = \sum_{l=1}^{\infty} [(l+1) M_{l+} + l M_{l-}] P^\prime_l (x) ,\\
\notag \mathcal{F}_3 & = \sum_{l=1}^{\infty} \left\{ [E_{l+}-M_{l+} ] P^{\prime \prime}_{l+1}(x) + [E_{l-}+M_{l-} ] P^{\prime \prime}_{l-1}(x) \right\},\\
\mathcal{F}_4 & = \sum_{l=2}^{\infty} [M_{l+} - E_{l+} - M_{l-}-E_{l-}] P^{\prime\prime}_l(x) ,
\label{eq:multipoleseries}
\end{align}
where $x=\cos(\theta)$, $P_l(x)$ is a Legendre polynomial of degree
$l$, $P^\prime_l(x)=\dv{P_l}{x}(x)$ is its first derivative, $P^{\prime\prime}_l(x)$ is the second derivative with respect to $x$ and $l$ is the orbital angular momentum of the outgoing pion-nucleon system. The subscript $\pm$ denotes the total angular momentum $j=l\pm 1/2$. Equation~\eqref{eq:multipoleseries} can be inverted:
\begin{align}
	\notag E_{l+} & = \int\limits_{-1}^{1}\frac{\dd x}{2(l+1)} \left[ P_l \mathcal{F}_1 -P_{l+1} \mathcal{F}_2 + \frac{l}{2l+1} (P_{l-1}-P_{l+1}) \mathcal{F}_3 + \frac{l+1}{2l+3} (P_l-P_{l+2}) \mathcal{F}_4 \right],\\
	\notag E_{l-} & = \int\limits_{-1}^{1}\frac{\dd x}{2 l} \left[ P_l \mathcal{F}_1 - P_{l-1} \mathcal{F}_2 - \frac{l+1}{2l+1} (P_{l-1}-P_{l+1}) \mathcal{F}_3 + \frac{l}{2l-1}(P_{l}-P_{l-2}) \mathcal{F}_4 \right],\\
	\notag M_{l+} & = \int\limits_{-1}^{1}\frac{\dd x}{2(l+1)} \left[ P_l \mathcal{F}_1 - P_{l+1} \mathcal{F}_2 - \frac{1}{2l+1} (P_{l-1} - P_{l+1}) \mathcal{F}_3 \right],\\
	M_{l-} & =\int\limits_{-1}^{1}\frac{\dd x}{2l} \left[ -P_l \mathcal{F}_1 + P_{l-1} \mathcal{F}_2 + \frac{1}{2l+1} (P_{l-1}- P_{l+1}) \mathcal{F}_3 \right]\,.
\end{align}
Here, we suppress the $x$-dependence of the Legendre polynomials $P_l$ for the sake of brevity. 

To calculate multipole amplitudes, we proceed as follows: first, we express the pion photoproduction amplitude in terms of the Ball amplitudes (Eq.~\eqref{eq:ballamplitudes}). Then, we rewrite the $B_i$'s in terms of $A_i$'s to obtain the representation of the amplitude in the minimal basis (Eq.~\eqref{eq:minimalbasis}). Finally, we use the coefficients $A_i$ to calculate $\mathcal{F}_1 - \mathcal{F}_4$. The relations between these representations are given in Appendix~\ref{sec:relationsofAisandBis}.
%
%

Next, we provide the expressions for the unpolarized differential cross section and linear polarization asymmetry. The differential cross section for pion photoproduction is given by
\begin{align}
\dv{\sigma}{\Omega} = \frac{1}{64 \uppi^2 s} \frac{\abs{\vec{q}}}{\abs{\vec{k}}} \abs{\overline{\mathcal{M}}}^2,
\end{align}
with the unpolarized squared matrix element $\abs{\overline{\mathcal{M}}}^2$ 
\begin{align}
	\abs{\overline{\mathcal{M}}}^2 = \frac{1}{4} \sum\limits_{\lambda=-1}^{1} \sum\limits_{s,s^\prime=-1}^{1} \abs{\epsilon_{\mu}(k,\lambda) \mathcal{M}^{\mu}(k,p,s,p^\prime,s^\prime,q)}^2,
\end{align}
where again $\lambda$ is the helicity of the photon, $s$ ($s^\prime$) is the spin of the incoming (outgoing) nucleon and the factor of 1/4 arises from averaging over helicity and spin of the incoming particles. 
The linearly polarized photon asymmetry $\Sigma$ is given by
\begin{align}
\Sigma = \frac{\dd\sigma_{\perp}-\dd\sigma_{\parallel}}{\dd\sigma_{\perp}+\dd\sigma_{\parallel}}\,,
\label{eq:asymmetry}
\end{align}
where $\dd \sigma_{\perp}$ and $\dd \sigma_{\parallel}$ refer to the differential cross section for photon polarizations perpendicular and parallel to the reaction plane, respectively.
%
%
%
%
\section{Effective Lagrangian and Power Counting} 
\label{sec:Lagrangian}
The calculation of observables in chiral perturbation theory is based on Feynman rules derived from the effective Lagrangian. In general, it contains an infinite number of terms with a rising number of derivatives and/or pion masses. At the maximal order we are working, the terms of the effective Lagrangian relevant for the calculation of 
pion photoproduction read
\begin{align}
	\mathcal{L}_{\text{eff}} = \mathcal{L}_{\pi\pi}^{(2)} + \mathcal{L}_{\pi\pi}^{(4)} + \mathcal{L}_{\pi N}^{(1)} + \mathcal{L}_{\pi N}^{(2)} + \mathcal{L}_{\pi N}^{(3)}+ \mathcal{L}_{\pi N \Delta}^{(1)} + \mathcal{L}_{\pi N \Delta}^{(2)} + \mathcal{L}_{\pi N \Delta}^{(3)} + \mathcal{L}_{\pi \Delta}^{(1)} + \mathcal{L}_{\pi \Delta}^{(2)},
\end{align}
where we do not distinguish between HB and covariant notation at this point.

In the mesonic sector, the building blocks are the pion field $U=u^2$ with 
\begin{align}
U(x)= 1 + \ci \frac{\boldsymbol{\tau}\cdot \boldsymbol{\pi}}{F}-\frac{\pi^2}{2 F^2} - \ci \upalpha \frac{\pi^2\, \boldsymbol{\tau} \cdot \boldsymbol{\pi}}{F^3} + \left(\upalpha-\tfrac{1}{8}\right) \frac{\pi^4}{F^4}+\order{\pi^5},
\label{eq:generalpionfield}
\end{align}
where $F$ is the pion decay constant in the chiral limit and $\upalpha$ is an arbitrary unphysical parameter from the parametrization of the pion field. The covariant derivative acting on the pion field $\nabla_\mu$  is defined as 
\begin{align}
	\nabla_\mu U= \partial_\mu U - \ci r_\mu U + \ci U l_\mu, \quad 
	l_\mu=v_\mu-a_\mu, \quad 
	r_\mu=v_\mu+a_\mu,
\end{align}
where $v_{\mu}=-e\nucleoncharge A_{\mu}=-e\frac{\identity+\paulimatrix_3}{2}A_{\mu}$ is the vector source 
with the electric charge $ e\approx\num{0.303} $ and the electromagnetic field $A_{\mu}$, and $a_{\mu}$ is the axial source. Furthermore, 
we introduce $\chi=\diag(M^2,M^2)$ and $\barepionmass$ is the pion mass to leading order in quark masses. We also introduce the field combinations 
\begin{align}
	\chi_{\pm} =u^{\dagger}\chi u^{\dagger} \pm u \chi^{\dagger} u,\quad 
	F_L^{\mu\nu} =\partial^\mu l^\nu-\partial^\nu l^\mu - \ci [l^\mu,l^\nu],\quad 
	F_R^{\mu\nu}=\partial^\mu r^\nu-\partial^\nu r^\mu- \ci [r^\mu,r^\nu].
\end{align}

The relevant parts of the leading and next-to-leading pionic Lagrangian read \cite{Gasser:1983yg}
\begin{align}
\notag \mathcal{L}_{\pi \pi}^{(2)} &= \frac{F^2}{4} \trace \left( \nabla_{\mu} U \nabla^{\mu}U^\dagger \right) + \frac{F^2}{4} \trace \left( \chi_+\right),\\
\notag \mathcal{L}_{\pi\pi}^{(4)}& =\frac{l_3}{16} \trace(\chi_+)^2 + \frac{l_4}{16} \left( 2 \trace(\nabla_{\mu}U (\nabla^{\mu} U)^{\dagger}) \trace(\chi_+) + \trace(2((\chi U^{\dagger})^2+(U \chi^{\dagger})^2) - 4 \chi^{\dagger} \chi - \chi_-^2) \right)\\
& + \ci \frac{l_6}{2} \trace \left( F_{R,\mu\nu} \nabla^{\mu}U (\nabla^{\nu} U)^{\dagger} + F_{L,\mu\nu} (\nabla^{\mu}U)^{\dagger} \nabla^{\nu} U \right).
\label{eq:Lpipi}
\end{align}

Introducing nucleons, we give the definitions of 
\begin{align}
	\notag D_\mu & =\partial_\mu - \Gamma_\mu, 
	\quad \Gamma_\mu  =\frac{1}{2} \left\{ u^\dagger (\partial_\mu - \ci r_\mu) u+ u (\partial_\mu-\ci l_\mu) u^\dagger \right\}, \quad 
	u_\mu  =\ci (u^\dagger(\partial_\mu-\ci r_\mu)u - u(\partial_\mu- \ci l_\mu)u^\dagger), \\
	F_{\mu\nu}^{\pm} & = u^\dagger F_{\mu\nu}^R u \pm u F_{\mu\nu}^L u^\dagger, \quad
	\tilde{F}_{\mu\nu}^{\pm}  = F_{\mu\nu}^{\pm} - \frac{1}{2} \operatorname{Tr}(F_{\mu\nu}^{\pm}).
\end{align}

The leading order pion-nucleon Lagrangian in the covariant approach is given by
\begin{align}
 \mathcal{L}_{\pi N}^{(1)}=\bar{\Psi}_N & \left( \ci \slashed{D} - m + \frac{\baregA}{2}\slashed{u}\gamma_5 \right)\Psi_N,
\end{align}
where $m$ and $g$ are the bare nucleon mass and axial coupling constant.
Other relevant parts of the covariant $\pi N$ Lagrangian read (see, e.g., Ref.~\cite{Fettes:2000gb})
\begin{align}
	\notag \mathcal{L}_{\pi N}^{(2)} = \bar{\Psi}_N & \left\{ c_1 \trace(\chi_+) + \frac{c_6}{8m_N} F_{\mu\nu}^+ \sigma^{\mu\nu} + \frac{c_7}{8m_N} \trace(F_{\mu\nu}^+) \sigma^{\mu\nu} \right\} \Psi_N,\\
	\notag \mathcal{L}^{(3)}_{\pi N}= \bar{\Psi}_N & \left\{\frac{d_6}{2m_N} ( \ci \left[D^\mu, \tilde{F}_{\mu\nu}^+\right] D^\nu+\text{h.c.}) + \frac{d_7}{2m_N} ( \ci \left[D^\mu, \trace(F_{\mu\nu}^+)\right] D^\nu+\text{h.c.}) + \frac{d_8}{2m_N} ( \ci \levicivita^{\mu\nu\alpha\beta} \trace(\tilde{F}_{\mu\nu}^+ u_\alpha) D_\beta + \text{h.c.}) \right.\\
	\notag &  + \frac{d_9}{2m_N} ( \ci \levicivita^{\mu\nu\alpha\beta} \trace(F_{\mu\nu}^+)  u_\alpha D_\beta + \text{h.c.}) + \frac{d_{16}}{2} \gamma^{\mu} \gamma_5 \trace(\chi_+)u_\mu + \ci \frac{d_{18}}{2} \gamma^\mu \gamma_5 \left[D_\mu, \chi_-\right] \\
	\notag & \left. - \frac{d_{20}}{8m_N^2} (\ci \gamma^\mu \gamma_5 \left[\tilde{F}_{\mu\nu}^+,u_\alpha \right] D^{\alpha \nu} +\text{h.c.}) + \ci \frac{d_{21}}{2} \gamma^{\mu}\gamma_5 \left[\tilde{F}_{\mu\nu}^+,u^{\nu}\right] +  \frac{d_{22}}{2} \gamma^\mu \gamma_5 \left[D^\nu, F_{\mu\nu}^-\right] \ \right\} \Psi_N, 
\end{align}
with $\sigma^{\mu\nu}=\frac{\ci}{2}[\gamma^\mu, \gamma^\nu]$. We use the convention $\levicivita^{0123}=-1$. 

In the heavy baryon formalism, the nucleon momentum is split according to
\begin{align}
	\label{eq:hbmomentumdecomposition}
	p_\mu = m_N v_\mu + P_\mu,
\end{align}
where the first part is a large piece close to the on-shell kinematics and the second part $P_\mu$ is a soft residual contribution $v \cdot P \ll \nucleonmass$. The vector $v_\mu$ is the four-velocity of the nucleon with the properties $v^2=1, v^0 \geq 1$ and can be conveniently chosen as $v=(1,0,0,0)$. The nucleon field $\Psi_N$ is split into the so-called light and heavy fields 
\begin{align}
	\label{eq:hbfields}
	N_v(x) = \operatorname{e}^{\ci m_N v\cdot x} P_v^+ \Psi_N(x) \qq{and} h_v(x)=\operatorname{e}^{\ci m_N v\cdot x} P_v^- \Psi_N(x),
\end{align}
which are eigenstates of $\slashed{v}$, with the projectors 
\begin{align}
	P_v^{\pm}=\frac{1}{2}(1 \pm \slashed{v}).
	\label{eq:HBspinprojectionmatrices}
\end{align}
In the heavy baryon formalism, any bilinear $\bar{\Psi}\Gamma\Psi$ with $\Gamma \in \{ 1, \gamma_5, \gamma_\mu, \gamma_5 \gamma_\mu, \sigma_{\mu\nu} \}$ can be expressed in terms of the velocity $v^\mu$ and the Pauli-Lubanski spin vector 
\begin{align}
	S_\mu = \frac{\ci}{2}\gamma_5\sigma_{\mu \nu} v^\nu\,.
\end{align}
The relevant terms of the Lagrangians up to third order read \cite{Fettes:2000gb}
\begin{align}
	\notag \hat{\mathcal{L}}_{\pi N}^{(1)}  = &\, \bar{N} \left( \ci v \cdot D + g\, S\cdot u\right) N\,, \\
	\notag \hat{\mathcal{L}}_{\pi N}^{(2)} = &\, \bar{N}
	\left(c_1 \trace(\chi_+)
	- \frac{\ci}{4m_N} \left[ \hat{c}_6 \left[ S^\mu,S^\nu \right] F_{\mu\nu}^+ + \hat{c}_7 \left[ S^\mu,S^\nu \right] \trace(F_{\mu\nu}^+)\right] \right) N\,, \\
	\notag \hat{\mathcal{L}}_{\pi N}^{(3)}  = &\, \bar{N} \left( \hat{d}_6 \left[D^\mu, \tilde{F}_{\mu\nu}^+ \right] v^\nu + \hat{d}_7 \left[D^\mu, \trace(F_{\mu\nu}^+) \right] v^\nu + \hat{d}_8 \trace(\tilde{F}_{\mu\nu}^+ u_\alpha) \levicivita^{\mu\nu\alpha\beta} v_\beta + \hat{d}_9 \trace(F_{\mu\nu}^+) u_\alpha \levicivita^{\mu\nu\alpha\beta} v_\beta \right.\\
	\notag& \left. \quad + \hat{d}_{16}\, S\cdot u \trace(\chi_+) + \ci \hat{d}_{18} \left[S\cdot D, \chi_-\right] + \ci \hat{d}_{20} S^\mu v^\nu \left[ \tilde{F}_{\mu\nu}^+, v\cdot u \right]  + \ci \hat{d}_{21} S^\mu \left[ \tilde{F}_{\mu\nu}^+, u^\nu \right] + \hat{d}_{22} S^\mu \left[D^\nu, F_{\mu\nu}^-\right]  \right) N\\
	&+ \frac{1}{2m_N} \bar{N} \left( (v\cdot D)^2 -D^2-\ci \baregA \left\{S\cdot D,v\cdot u \right\} \right) N\,.
	\label{eq:HBLagrangians}
\end{align}
The prefactor of $1/m_N$ in the $c_i$ parts in $\hat{\mathcal{L}}_{\pi N}^{(2)}$ should not be interpreted as a $1/m_N$ correction; it originates from the definition of the constants $\hat{c}_6$ and $\hat{c}_7$, which are dimensionless when defined this way. Unlike the usual treatment of the single nucleon sector in the literature, we from now on consider the $1/m_N$ corrections as $1/m_N\sim 1/\Lambda^2$, 
where $\Lambda$ is the hard scale (see below),
so that they start to appear at the third instead of the second order. 
The $1/m_N^2$ are, therefore, shifted beyond the order we are working.
This is a common practice in studies of the nuclear forces \cite{Epelbaum:2008ga} and is referred to as $NN$ counting in this work. Also, note that the heavy baryon LECs $\hat{c}_6$ and $\hat{c}_7$ are not the same as the covariant constants, but they 
absorb $1/m_N$-shifts from the leading order Lagrangian via $\hat{c}_6 = c_6 + 1, \hat{c}_7  =c_7$.

The $\deltapart$ resonance is introduced as an explicit degree of freedom by an isospin-$3/2$ Rarita-Schwinger spinor $\bar{\Psi}_{\Delta,i}^\mu$, which satisfies $\tau_i \Psi_{\Delta,\mu}^i=0$ \cite{Hemmert:1997ye}. Given the following definitions:
\begin{align}
	D^{ij}_\mu =\partial_\mu \delta^{ij} + \Gamma^{ij}_\mu,
	\quad \Gamma^{ij}_\mu = \delta^{ij} \Gamma_\mu - \ci \levicivita^{ijk} \trace(\tau_k \Gamma^{\mu}),
	\quad u^{ij}_\mu  = \xi^{ik}_{3/2}\,u_\mu\,\xi^{kj}_{3/2}
\end{align}
with the isospin projectors
\begin{align}
	\xi_{ij}^{3/2} = \frac{2}{3} \delta_{ij} - \frac{\ci}{3} \levicivita_{ijk} \tau_k, \quad \xi_{ij}^{1/2} = \frac{1}{3} \delta_{ij} + \frac{\ci}{3} \levicivita_{ijk} \tau_k, 
\end{align}
the leading covariant $\pi\Delta$-Lagrangian reads \cite{Hemmert:1997ye, PhDHemmert}
\begin{align}
	\notag \mathcal{L}^{(1)}_{\pi\Delta}=&- \bar{\Psi}_{\Delta,i}^\mu \left\{ (\ci \slashed{D}^{ij} - \mathring{m}_\Delta \delta^{ij}) g_{\mu\nu} -\ci (\gamma_\mu D_\nu^{ij}+ \gamma_\nu D_\mu^{ij}) + \ci\gamma_\mu \slashed{D}^{ij}\gamma_\nu + \mathring{m}_\Delta \gamma_\mu \gamma_\nu \delta^{ij} \right. \\
	& \qquad \qquad \left. + \frac{\baregOne}{2} g_{\mu\nu} \slashed{u}^{ij} \gamma_5 + \frac{\mathring{g}_2}{2} (\gamma_\mu u_\nu^{ij} + u_\mu^{ij} \gamma_\nu) \gamma_5 + \frac{\mathring{g}_3}{2} \gamma_\mu \slashed{u}^{ij} \gamma_5 \gamma_\nu \right\} \Psi_{\Delta,j}^\nu\,,
	\label{eq:LpiDeltaDelta}
\end{align}
where $\mathring{m}_\Delta$ is the bare $\deltapart$ mass and 
$\baregOne$, $\mathring{g}_2$, $\mathring{g}_3$ are the bare leading order coupling constants.
For pion photoproduction, only $\baregOne$ is relevant, whereas the constants $\mathring{g}_2$ and $\mathring{g}_3$ are off-shell parameters, which do not contribute if the $\deltapart$-particle is on-shell and result only in shifts of LECs. 

The only term from the second order $\pi\Delta$-Lagrangian relevant for our calculation is
\begin{align}
 \mathcal{L}^{(2)}_{\pi\Delta}=
 -\ci c_1^{\deltapart} \bar{\deltafield}_{\deltapart,\mu}^i  \trfl{\chi_+}\tensorsigma^{\mu\nu} \deltafield_{\deltapart,\nu}^i\,,
\end{align}
which enters barely the renormalization of the $\deltapart$ mass.

We also need to take into account the nucleon-to-$\deltapart$ transition Lagrangian $\mathcal{L}_{\pi N \Delta}$ up to third order. Again, we only give the relevant terms for pion photoproduction up to our working order (\cite{Hemmert:1997ye, PhDZoller}):
\begin{align}
\notag \mathcal{L}_{\pi N \Delta}^{(1)}  = &\, \barehA \left( \bar{\Psi}_{\Delta,i}^{\mu} \Theta_{\mu\nu}(z_0) w_i^\nu \Psi_N + \bar{\Psi}_N w^{\nu\dagger}_i \Theta_{\nu\mu}(z_0) \Psi_{\Delta,i}^\mu\right),\\
\notag \mathcal{L}_{\pi N \Delta}^{(2)}  = &\, \ci \frac{b_1}{2}  \bar{\Psi}_{\Delta,i}^{\mu} \Theta_{\mu\nu}(z_1)  F_i^{+,\nu\alpha} \gamma_\alpha \gamma_5 \Psi_N + \ci b_3 \bar{\Psi}_{\Delta,i}^{\mu}\Theta_{\mu\nu}(z_3) w_i^{\nu\alpha}\gamma_{\alpha} \Psi_N - \frac{b_6}{m} \bar{\Psi}_{\Delta,i}^\mu \Theta_{\mu\nu}(z_6)  w_i^{\nu\alpha}  D_{\alpha} \Psi_N + \dots + \text{h.c.},\\
\notag \mathcal{L}_{\pi N \Delta}^{(3)} = &\, \frac{h_1}{m} \bar{\Psi}_{\Delta,i}^{\mu} \Theta_{\mu\nu}(y_1) F_i^{+,\nu\alpha} \gamma_5 D_\alpha \Psi_N - \ci \frac{h_{15}}{2} \bar{\Psi}_{\Delta,i}^\mu \Theta_{\mu\nu} (y_{15}) \operatorname{Tr}\left( [D^\alpha,F^{+,\nu\beta}] \tau^i \right)\sigma_{\alpha \beta} \gamma_5 \Psi_N\\
& + \ci \frac{h_{16}}{2m} \bar{\Psi}_{\Delta, i}^\mu \Theta_{\mu\nu}(y_{16}) \operatorname{Tr} \left( [D^{\alpha}, F^{+,\nu\beta}] \tau^i \right) \gamma_\beta \gamma_5 D_\alpha \Psi_N + \dots + \text{h.c.}\, ,
\label{eq:LpiNDelta} 
\end{align}
where $\barehA$, $b_i$ and $h_i$ are LECs and 
%
\begin{align}
	w_i^\mu=\frac{1}{2} \operatorname{Tr}\big(\tau_i u^\mu\big), \quad w_i^{\mu\nu}=\frac{1}{2} \operatorname{Tr}\big(\tau_i [D^\mu, u^\nu]\big), \quad F_{i,\mu\nu}^{\pm}=\frac{1}{2}\operatorname{Tr}\big( \tau_i F_{\mu\nu}^{\pm} \big), \quad 
	\Theta_{\mu\nu} (z)= g_{\mu\nu} + z \gamma_\mu \gamma_\nu.
\end{align}
$\Theta_{\mu\nu}(z)$ is an off-shell function with the off-shell parameter $z$. It was shown in Ref.~\cite{Krebs:2009bf} that the dependence on the off-shell parameters can be eliminated by a redefinition of the LECs up to higher order corrections. Thus, we set them to zero in our calculations.
The $\deltapart$ propagator \cite{Bernard:2003xf} derived from the Lagrangian~\eqref{eq:LpiDeltaDelta} reads 
\begin{align}
	\label{eq:deltapropagator}
	\mathcal{G}_{\Delta, ij}^{\mu\nu} (p) = - \frac{\ci (\slashed{p} + \baredeltamass)}{p^2 - \baredeltamass^2} \left( g^{\mu\nu} - \frac{1}{d-1} \gamma^{\mu} \gamma^{\nu} + \frac{1}{d-1} \frac{p^\mu \gamma^{\nu} - p^{\nu} \gamma^{\mu} }{\baredeltamass} + \frac{d-2}{d-1} \frac{p^\mu p^\nu}{\baredeltamass^2} \right) \xi_{ij}^{3/2}
\end{align}
with the isospin projector 
\begin{align}
	\xi_{ij}^{3/2} = \frac{2}{3} \delta_{ij} - \frac{\ci}{3} \levicivita_{ijk} \tau_k.
\end{align}

%

%
%
We also give our power counting scheme. In the $\deltapart$-less case, we employ
\begin{align}
	\label{eq:nucleoniccountingrules}
	Q=\frac{q}{\Lambda} \in \left\{  \frac{M_\pi}{\Lambda}, \frac{k}{\Lambda} \right\} \qq{with} \Lambda \in \left\{ \Lambda_b, 4\pi F_\pi \right\},
\end{align}
where $k$ is an arbitrary three-momentum (of a meson or a baryon),
$F_\pi$ is the pion decay constant and $\Lambda_b$ is the chiral symmetry breaking scale.
In the $\deltapart$-full theory, we introduce the $\deltapart$-nucleon-mass split $\Delta=\deltamass-\nucleonmass$,
which we count as an additional small parameter of order $\order{M_\pi}$, but which does not vanish in the chiral limit. 
In the $\deltapart$-full case, we use the small scale expansion (SSE) counting scheme \cite{Hemmert:1997ye}
\begin{align}
	\label{eq:deltacountingrules}
	\epsilon \in \left\{ \frac{M_\pi}{\Lambda}, \frac{k}{\Lambda}, \frac{\Delta}{\Lambda} \right\} \qq{with} \Lambda \in \left\{ \Lambda_b, 4\pi F_\pi \right\},
\end{align}
with $\epsilon$ as the new expansion parameter in the $\deltapart$-full case in contrast to $Q$ in the $\deltapart$-less theory.

The order $D$ of any Feynman diagram can be computed according to the formula \cite{Weinberg:1991um}
\begin{align}
D=1+2L+\sum_{n}(2n-2)V_{2n}^M+\sum_d(d-1)V_d^B\,,
\label{eq:power_counting_D}
\end{align}
where  $ L $ is the number of loops, $ V^M_{2n} $ is the number of purely mesonic vertices of order $ 2n $ and $ V^B_d $ is the number of vertices 
involving baryons of order $d$. 

As mentioned above,
the $1/m_N$ corrections in the heavy-baryon scheme are treated following the $NN$ counting, so that
$ \frac{q}{m_N}\sim  \mathcal{O}\left( Q^2 \right)$.

%
%
%
%
\section{Renormalization} 
\label{sec:renormalization}
In this section, we discuss the subtleties of renormalization, i.e.\ all steps necessary to remove unphysical infinities
and possible power counting breaking terms from the theory. 
\subsection{Renormalization of subprocesses}
\label{sec:renormalizationsubprocesses}
To relate the bare constants of the effective Lagrangian to their physical counterparts, we must consider various subprocesses of pion photoproduction. We take care of the appearing integrals and their ultraviolet (UV) divergences by dimensional regularization. Also, we choose to renormalize the masses, wave functions and coupling constants using the on-shell renormalization. 
Our renormalization scheme and the renormalization conditions for the pion, nucleon and $\deltapart$ self-energies,
pion decay constant, $\pi NN$, $\pi N\deltapart$ and $\pi \deltapart\deltapart$ coupling constants,
the nucleon magnetic moments and the electromagnetic $N\deltapart$ transition form factor
were described in detail in Ref.~\cite{Rijneveen:2020qbc}. 
For completeness, we provide the expressions for all counterterms needed in our calculation
in Appendix~\ref{sec:counterterms}.
Note that loops with internal $\deltapart$-lines appearing at order $\epsilon^3$ were not included in 
Ref.~\cite{Rijneveen:2020qbc}. The corresponding additional contribution to the counter terms are indicated
by the superscript $\deltapart$, e.g., $\delta \nucleonmass^{(3,\Delta)}$.
For the $\deltapart$-mass renormalization we use the so-called complex mass scheme
\cite{Denner:1999gp,Denner:2005fg,Denner:2006ic,Yao:2016vbz}
in exactly the same manner as in Ref.~\cite{Rijneveen:2020qbc}.

\subsection{Renormalization of the photoproduction LECs}
\label{sec:renormalizationofLECs}
Additionally to the UV divergences removed by on-shell renormalization of the subprocesses, 
it is necessary to perform a renormalization of the photoproduction LECs to absorb the remaining
divergences and (possibly) power counting breaking terms.
In the covariant approach, we employ the EOMS renormalization scheme \cite{Fuchs:2003qc}.
The LEC shifts are defined as
\begin{align}
	d_i=\bar{d_i} + \beta_{d_i}\frac{\bar{\lambda}}{\piondecayconstant^2}\\
	\intertext{in the covariant approach and as}
	\hat{d}_i = \hat{d}_i^r + \beta_{\hat{d}_i}\frac{\bar{\lambda}}{\piondecayconstant^2}
\end{align}
in the HB approach, 
where
\begin{align}
	\label{eq:lambdabar}
	\bar{\lambda} = \frac{1}{16 \uppi^2} \left[\frac{1}{d-4}+\frac{1}{2}(-1+\gamma_E-\ln 4\uppi)\right]\,,
\end{align}
and $d$ is the space-time dimension.
The renormalized LECs are denoted by the superscript “$r$” in the HB approach and by the bar in the covariant case.
Note that in this work, we use a more standard definition of $\bar d_i$ as compared to Ref.~\cite{Rijneveen:2020qbc}.

The renormalization scale $\mu$ in all integrals (see Appendix~\ref{sec:loop_integrals})
is set to $\mu=\nucleonmass$ ($\mu=\pionmass$) for the covariant (heavy-baryon) scheme
to be consistent with traditional choices in the literature.

In the $\deltapart$-less HB sector, the beta functions were given in Ref.~\cite{Gasser:2002am}:
\begin{align}
	\beta_{\hat{d}_8} = 0, \quad \beta_{\hat{d}_9} = 0, \quad \beta_{\hat{d}_{20}} = \axialcoupling + \axialcoupling^3, \quad \beta_{\hat{d}_{21}}=-\axialcoupling^3, \quad \beta_{\hat{d}_{22}}=0.
\end{align}
In the covariant approach, they can easily be derived by expanding the
UV divergent piece of the whole scattering amplitude in the small
scales up to the working order. In the $\deltapart$-less approach,
they are found to vanish,
\begin{align}
	\beta_{d_8} = \beta_{d_9} = \beta_{d_{20}} = \beta_{d_{21}} = \beta_{d_{22}} =0,
\end{align}
but for the full $\epsilon^3$ amplitude, they read
\begin{gather}
	\notag \beta_{d_8} = \frac{19}{54}\axialcoupling \pindcoupling^2- \frac{55}{729} g_1 \pindcoupling^2, \hspace{0.3cm} \beta_{d_9} =-\frac{97}{324}\axialcoupling \pindcoupling^2+ \frac{55}{1458} g_1 \pindcoupling^2,\hspace{0.3cm} \beta_{d_{20}} = \frac{47}{162}\axialcoupling \pindcoupling^2-\frac{2555}{1458} g_1 \pindcoupling^2,\\[0.3cm]
	\beta_{d_{21}} = -\frac{823}{162} \axialcoupling \pindcoupling^2+ \frac{10}{27} g_1 \pindcoupling^2, \hspace{0.3cm} \beta_{d_{22}} =\frac{40}{27} \axialcoupling \pindcoupling^2-\frac{425}{243} g_1 \pindcoupling^2.
\end{gather}
Here and in what follows, $\axialcoupling$ and $\pindcoupling$ are 
the nucleon axial coupling  and the $\pi N \deltapart$ coupling constant, respectively.  
In the EOMS scheme, the power-counting violating terms must be subtracted by shifting the LECs. 
Up to the order we are working, it can easily be argued that power-counting violating terms cannot occur,
since the symmetry constraints allow contact interactions in the photoproduction sector starting only 
from order ${Q}^3$.
We explicitly checked the absence of power-counting violating terms by replacing all occurring integrals in the ${Q}^3$ amplitude by their infrared regular part and performing an explicit expansion in the small scales. We verified that all power-counting violating terms cancel out, which provides a valuable consistency check, because nontrivial cancellations are analytically fulfilled. In the $\deltapart$-full approach at order $\epsilon^3$, we refrained from repeating this exercise, since the same argument as in the $\deltapart$-less case holds and performing the comparable test would require considerably more effort. We conclude that for our study, the EOMS scheme is effectively equivalent to the $\widetilde{MS}$ scheme \cite{Gasser:1983yg, Fuchs:2003qc}.

As explained in Ref.~\cite{Rijneveen:2020qbc}, the constants $b_3,
b_6, h_{15}$ and $h_{16}$ are redundant at our working order. The
shifts to absorb these LECs in the covariant order-$\epsilon^3$
amplitude through a redefinition of $\pindcoupling$, $b_1$ and $h_1$
are given by $d_i \to d_i + \delta_{d_i}$ with
\begin{align}
	\delta_{d_8}= \delta_{d_{20}} = -\delta_{d_{21}}= \frac{-b_1(b_3+b_6) + 2 \pindcoupling(h_{15}+h_{16})}{9}, \quad \delta_{d_9} = \delta_{d_{22}} =0.
\end{align}

%
%
%
%
\section{Results and discussion} 
\label{sec:results} 
In this section, we present the results of our calculation. We used our own code written in \textit{Mathematica} \cite{Mathematica}, \textsc{FORM} \cite{Kuipers:2012rf} and Fortran and relied on LoopTools \cite{Hahn:1999} and on the \textit{X} package \cite{Patel:2015tea} for the numerical evaluation of loop integrals.
\subsection{Low-energy constants}
\label{sec:LECs}
In the $\deltapart$-less theory, there are four independent parameters
to be determined, which are $d_8, d_9, d_{20}$ and
$d_{21;22}=d_{21}-d_{22}/2$. Note that while the numerical values of LECs are, in general, different in the HB and covariant approaches, the fitting procedure is similar. Thus, when we discuss points applicable to both formalisms in the $\deltapart$-less fits, we simply use $d_i$ for referring to both renormalized parameters $\bar{d}_i$ and $\hat{d}^{r}_i$. Note that one cannot assess the constants $d_{21}$ and $d_{22}$ individually in real pion photoproduction, we can determine only the combination $d_{21;22}$. When including the leading-order $\deltapart$ tree diagrams of order $\epsilon^2$, the additional constant $b_1$ enters the amplitude. $b_1$ corresponds to the leading-order $\gamma N \deltapart$ coupling constant. At order $\epsilon^3$, the additional coupling $h_1$, the subleading $\gamma N \deltapart$ coupling is introduced. The values of the constants taken from other sources are collected in Table~\ref{tab:LO_constants}. The value of the $\deltapart$ mass was determined from a fit to the order-$\epsilon^2$ amplitude as described in Ref.~\cite{Rijneveen:2020qbc}.
\begin{table}[htbp]
	\caption{Particle masses (in MeV) and leading-order coupling constants used in this work.
		Unless specified, the values are taken from
		PDG~\cite{Zyla:2020zbs}.
	} \label{tab:LO_constants}
	\begin{tabular*}{\textwidth}{@{\extracolsep{\fill}}cccccccc}
		\hline
		&&&&&&& \\[-7pt]
		$ \pionmass$ & $ \nucleonmass $ & $\deltamass$ (fit) & $ e $ & $
		\piondecayconstant
		\; [\si{\MeV}] $ & $\axialcoupling$ & $\pindcoupling$ & $g_1$\\[2pt]
		\hline
		&&&&&&& \\[-7pt] 
		\num{138.03} & \num{938.27} & \num{1219.3} - \num{53.7}\,\ci & \num{0.303}
		& \num{92.1} & \num{1.289} \cite{Baru:2010xn} & \num{1.43}
		\cite{Bernard:2012hb,Yao:2016vbz}&
		\num{-1.21} \cite{Yao:2016vbz}\\[2pt]
		\hline
	\end{tabular*}
\end{table}

The numerical values of LECs in the covariant and in the heavy baryon
approach are, in general, not directly comparable. First, the values
are related by $1/\nucleonmass$ corrections, which originate from the
construction of the HB Lagrangian, where a strict expansion in terms
of the inverse nucleon mass is done, such that the dependence on
$\nucleonmass$ is shifted to a series of additional contact
interactions suppressed by powers in $1/\nucleonmass$. Note that the
corrections are only relevant if the working order is beyond the order
at which the LECs appear first. For example, the values of $b_1$ in
both frameworks can be compared directly if only the leading
$\deltapart$ tree contributions (order $\epsilon^2$) are included,
because the difference starts to arise from
$1/\nucleonmass$-corrections, which, however, are beyond the working
order. For LECs appearing at the loop level, an extra shift generated
by the infrared regular (IR) part of the loop integrals must be
considered (the way of calculating the IR part of a loop integral is
described in e.g.\ Ref.~\cite{Fuchs:2003qc}). In the covariant
framework, the IR part gives rise to additional numerical
contributions, which affect the fitted values of the LECs. Because
these parts are absent in the HB approach, the covariant LECs cannot
be compared with the HB values, even if inverse nucleon mass
corrections are irrelevant. However, the contributions of the IR part
to the covariant LECs can be calculated analytically and switched off
in order to restore the same meaning of the LECs in the HB and
covariant approaches.
We determined these IR shifts by replacing the integrals of the covariant amplitude by their corresponding IR term and adjusting the constants in such a way that the obtained expression vanishes. The shifts we found for the relevant $d_i$'s  are given by
\begin{align}
	\notag \Delta d_8^{\text{IR}} & =\bar d_8 - \hat{d}_8 = \frac{\axialcoupling}{128 \piondecayconstant^2 \uppi^2}(3+\axialcoupling^2),\\
	\notag \Delta d_9^{\text{IR}} & = \bar d_9 - \hat{d}_9 = \frac{\axialcoupling}{128 \piondecayconstant^2 \uppi^2} (-1+\axialcoupling^2),\\
	\notag \Delta d_{20}^{\text{IR}}&  =\bar d_{20} -\hat{d}_{20}=  \frac{\axialcoupling}{96 \piondecayconstant^2 \uppi^2} 
	\left[6+11 \axialcoupling^2+3(1+\axialcoupling^2)\left(1- \gamma_E+ \ln(4\uppi)+\ln(\pionmass^2/\nucleonmass^2)\right)\right] , \\
	\Delta d_{21;22}^{\text{IR}} & = \bar d_{21;22}- \hat{d}_{21;22} = -\frac{\axialcoupling}{96 \piondecayconstant^2 \uppi^2} \left[3+8\axialcoupling^2+3\axialcoupling^2 \left(1- \gamma_E+ \ln(4\uppi)+\ln(\pionmass^2/\nucleonmass^2)\right)\right],
	\label{eq:IRshifts}
\end{align}
where we recall that the constants with the hat refer to HB parameters. 

The numerical differences between LECs determined in a
$\deltapart$-less and a $\deltapart$-full framework are expected to be
qualitatively described in terms of resonance saturation by
calculating the dominating leading-order contributions of the
$\deltapart$ to specific LECs, as stated in the decoupling theorem
\cite{Appelquist:1974tg}. In our case, this is accomplished by
expanding the order-$\epsilon^2$ $\deltapart$-full tree diagrams in
inverse powers of $\Delta \equiv \deltamass-\nucleonmass$
and matching the obtained expressions with the contact term structures of the $\deltapart$-less theory. This procedure yields
\begin{align}
	\delta d_8(\Delta) = -\delta d_{21;22}(\Delta) = -\frac{\pindcoupling b_1}{9\Delta}, \quad \delta d_9(\Delta) = \delta d_{20}(\Delta)=0,
\end{align}
with $\delta d_i(\Delta)= \bar{d}_i^{\slashed{\Delta}}-\bar{d}_i^{\Delta}$ denoting the difference between the constant $\bar{d}_i$ in the $\deltapart$-less (superscript $\slashed{\Delta}$) and $\deltapart$-full (superscript $\Delta$) approaches.
\subsection{Fitting procedure and Bayesian uncertainties}
Ideally, one would fit the LECs to the whole set of available pion photoproduction data in the relevant energy range. However, analyzing all the data requires a lot of effort and is an art of its own, since there is a lot of data available on the photoproduction process. This is why we choose to fit to the multipoles of the partial-wave analysis (PWA) from Mainz, the MAID2007 model \cite{Drechsel:2007if}, which is a more pragmatic approach. Furthermore, we only fit to the real part of the multipoles, because the imaginary part follows from unitarity as stated by Watson's theorem \cite{Watson:1954uc} and does not provide new information. We fit to the multipoles in the isospin channels, which is a natural choice because isospin-breaking effects are not considered in this work. The four physical reaction channels are linear combinations of the isospin channels. In the following, we remark on several points relevant for the fits.

\subsubsection{Uncertainties and fitting procedure}
The main disadvantage of using the MAID PWA is that uncertainties are not provided. We thus assign a relative $\SI{5}{\percent}$ error to every multipole, which is a common approach, see e.g.\ Ref.~\cite{Fettes:1998ud} for a similar procedure in the analysis of pion-nucleon elastic scattering, such that the uncertainty of the observables reads for our case
\begin{align}
\delta O_i = \sqrt{\Big(0.05\, O_i^{\text{exp}}\Big)^2+\Big(\delta O_i^{(n)}\Big)^2}
\label{eq:combineduncertainty}
\end{align}
with $O_i^{\text{exp}}$ the given value of the observable and $\delta O_i^{(n)}$ the truncation error at order $n$.
To estimate the uncertainties originating from the truncation of the chiral expansion 
we follow the Bayesian approach described in Ref.~\cite{Rijneveen:2020qbc}, which is based
on the previous developments in the literature, see Refs.~\cite{Furnstahl:2015rha, Melendez:2017phj, Epelbaum:2019zqc}. 
As compared to Ref.~\cite{Rijneveen:2020qbc}, we adopt a more appropriate 
in the various considered energy regions
definition of the small parameter $Q$, following Ref.~\cite{Epelbaum:2019zqc}:
\begin{align}
 Q= \max \bigg( \frac{E_\pi}{\Lambda_b}, \;
  \frac{M_\pi^\text{eff}}{\Lambda_b} \bigg) \,,
\end{align}
with $M_\pi^\text{eff} = 200$~MeV, instead of $Q= E_\pi/\Lambda_b$.

We have checked that the choice of the relative ``experimental'' error has negligible effects on the fit result by varying its size between $\SI{2}{\percent}$ and $\SI{15}{\percent}$. Of course, our choice affects the fit quality, but we found that these effects are relatively small, because the uncertainties are dominated by the Bayesian truncation errors.
To obtain the central values of the fit parameters, we minimize the $\chi^2$ function
\begin{align}
\chi^2=\sum\limits_{i}\left( \frac{O_i^{\text{exp}}-O_i^{(n)}}{\delta O_i} \right)^2, 
\end{align}
where the sum runs over all energy points from every multipole we incorporate in the configuration. 

After obtaining the central values of the fit parameters, the corresponding uncertainties are determined from the covariance matrix, which is approximated by the inverse of the Hessian
\begin{align}
\delta y_i = \sqrt{\text{Cov}(y_i,y_i)},\quad  \text{Cov}(y_i,y_j)= H_{ij}^{-1}, \quad H_{ij} = \frac{1}{2} \pdv{\chi^2}{y_i}{y_j} \bigg|_{\boldsymbol{y}=\bar{\boldsymbol{y}}}.
\label{eq:covariancematrix}
\end{align}
Here, the vector $\boldsymbol{y}$ refers to the set of all fitting parameters, and $\bar{\boldsymbol{y}}$ is the vector of best fit values.

\subsubsection{Energy range}
The energy range in which $\chi$PT is applicable is limited by the lowest lying not included resonance, which is in the $\deltapart$-less formulation the $\deltapart$ particle. Thus, the $\deltapart$ energy region should be excluded completely in the $\deltapart$-less fit, so that we choose to restrict the upper energy boundary to $\sqrt{s} = \SI{1200}{\mega\electronvolt}$. By construction, we expect the range of convergence of HB$\chi$PT to be smaller than in the covariant case. Also, in previous studies, it was shown that the HB approach yields good agreement with the neutral pion production data only up to $\SI{20}{\mega\electronvolt}$ above threshold \cite{FernandezRamirez:2012nw}. Furthermore, we exclude the region very close to the pion production threshold. Since we work in the isospin-symmetric limit, which leads to equal pion masses, our calculation cannot account for effects generated by the mass difference of the pions. The reaction threshold lies at $\sqrt{s}\approx\SI{1076}{\mega\electronvolt}$, so we restrict the lower energy boundary of our fits to $\sqrt{s}=\SI{1090}{\mega\electronvolt}$. We choose energy steps of $\SI{2}{\mega\electronvolt}$, so we have 56 data points per observable in the $\deltapart$-less case.	Briefly stated, we use the energy range  $\SI{1090}{\mega \electronvolt} \leq \sqrt{s} \leq \SI{1200}{\mega \electronvolt}$
in the $\deltapart$-less fit.	

In the $\deltapart$-full case, we extend the upper energy boundary to $\sqrt{s} = \SI{1250}{\mega\electronvolt}$ to partially take into account the $\deltapart$ region. By construction of the complex-mass approach, unitarity is strongly violated
very close to threshold due to the constant imaginary part of the
$\deltapart$ mass in the $\deltapart$ pole diagrams. However,
unitarity will be restored perturbatively when including higher
orders. As long as we first include only the $\deltapart$ tree
diagrams, we expect the description of the data to be worse near
threshold compared to the $\deltapart$-less approach. However, these
effects should be dominant in the imaginary parts of the multipoles,
which we do not fit. Therefore, we still choose to include the threshold region in our fitting range and comment on the description of the threshold region in the next section. Stated briefly, we fit between $\SI{1090}{\mega \electronvolt} \leq \sqrt{s} \leq \SI{1250}{\mega\electronvolt}$.

In our $\epsilon^3$ study, we have again modified the energy range. In
principle, the same arguments as in the case of ${Q}^3+\epsilon^2$ hold,
but we found that the data in the threshold region significant affect
the subleading $\gamma N \deltapart$ coupling constant $h_1$, while
having little impact
on the other LECs. Because we are interested in a precise determination of the $\gamma N \deltapart$ couplings, we removed the threshold region in the $\epsilon^3$ fits and take into account only the data between $\SI{1150}{\mega \electronvolt} \leq \sqrt{s} \leq \SI{1250}{\mega\electronvolt}$.

\subsubsection{Data configuration}
We restrict ourselves to the analysis of $s$- and $p$-wave multipoles, because they contain by far the largest contributions to the photoproduction cross sections.
Moreover, in higher partial waves, the unknown LECs contribute only as $1/m_N$ corrections.
In Fig.~\ref{fig:HBqTo3plot}, the results of the MAID analysis are depicted \cite{Drechsel:2007if}. Furthermore, the results of the energy-dependent \cite{Briscoe:2019cyo, PCStrakovsky} and energy-independent \cite{Workman:2012jf} GWU-SAID multipole amplitudes are shown in order to illustrate differences between various partial-wave analyses. The agreement between the three sets of data is excellent for the $M_{1+}^{3/2}$ multipole, which corresponds to the magnetic excitation of the $\deltapart$ resonance in the $s$-channel, and reasonably good for the $E_{0+}$ and $M_{1-}$ multipoles. 

In the $\deltapart$-less case, we are interested in choosing a fit
configuration sensitive to the values of the LECs $d_8, d_9, d_{20}$
and $d_{21;22}$. Therefore, it is instructive to analyze the
contributions of the parameters to the various amplitudes. The three
multipoles $E_{0+}, M_{1+}$ and $M_{1-}$ all receive leading-order
contributions of one or several LECs with respect to the $1/\nucleonmass$ expansion, i.e. $\mathcal{O}(m_N^0)$.
The multipole $E_{1+}$ gets only next-to-leading-order $1/\nucleonmass$ contributions from all four parameters, therefore we exclude $E_{1+}$ completely from the $\deltapart$-less fit. We choose to fit the $I=3/2$ channel first, which is motivated by the agreement of the data sets, the order of magnitude of the multipoles and bearing in mind that we are interested in analyzing the differences to the $\deltapart$-full theory, which is expected to bring dominant contributions to the $I=3/2$ channel. Altogether, we first fit to the three multipole amplitudes $E_{0+}^{3/2}$, $M_{1+}^{3/2}$ and $M_{1-}^{3/2}$, which determines the parameters $d_8, d_{20}$ and $d_{21;22}$. The constant $d_9$ does not contribute to the $I=3/2$ channel and is fitted to the $I=1/2$ channels subsequently. We anticipate that the constants $d_8, d_{20}$ and $d_{21;22}$ are sufficiently constrained from the $I=3/2$ fit to serve as an input for the $I=1/2$ fit. At leading order in $1/\nucleonmass$, $d_9$ contributes to both proton and neutron channels of $M_{1+}^{1/2}$ and $M_{1-}^{1/2}$. However, while performing the fit, we found that there is no value of $d_9$ which results in an acceptable description of $M_{1-}^{1/2}$. We expect that this problem is resolved if higher-order contributions are taken into account. Thus, we decided to exclude $M_{1-}^{1/2}$ from the second fit and determine $d_9$ only from $M_{1+}^{1/2}$. When removing $M_{1-}$, the central value of $d_9$ only changes very slightly, which supports our strategy, but of course the fit quality is affected. The uncertainty of $d_9$ is determined analogously to Eq.~\eqref{eq:covariancematrix} by 
\begin{align}
\delta d_9 = \left( \frac{1}{2} \pdv[2]{\chi_{I=1/2}^2}{d_9} \right)^{1/2}.
\end{align}
This procedure neglects the fact that $\delta d_9$ also receives
contributions from the uncertainties of the previously determined
LECs. We have checked that these effects indeed have a small impact,
so that it is legitimate to ignore them. This insight also supports our idea that $d_8$, $d_{20}$, $d_{21;22}$ are sufficiently well constrained by the first fit.

In the case of including the $\deltapart$-full $\epsilon^2$ tree
diagrams, the additional constant $b_1$ is introduced. In the
$s$-channel, $b_1$ contributes at leading order in $1/\nucleonmass$
only to $M_{1+}$, which is already part of our fitting
configuration. As $b_1$ contributes to the $I=3/2$ channel, we obtain
its central value as well as $\bar{d}_8$, $\bar{d}_{20}$ and
$\bar{d}_{21;22}$ from the $I=3/2$ fit. Subsequently, we determine
$\bar{d}_9$ from the $I=1/2$ channel as before. Note that at this
working order, it is not necessary to
distinguish
between covariant or HB $b_1$, because $1/\nucleonmass$ corrections and renormalization counterterms are beyond the working order. 

At order $\epsilon^3$, one includes the leading $\deltapart$-full loops
and the tree diagrams with the additional constant $\bar{h}_1$. 
The  $s$-channel delta-pole graphs with $\bar{h}_1$ give essential
contributions to the electric $\deltapart$ multipole
$E_{1+}$. Therefore, we slightly modify our fitting procedure and
include $E_{1+}$ in the $I=3/2$ fit, which determines in the following
the covariant LECs $\bar{d}_8$, $\bar{d}_{20}$, $\bar{d}_{21;22}$,
$\bar{b}_1$ and $\bar{h}_1$. Subsequently, we fit $\bar{d}_9$ to the
$I=1/2$ channel as before. At this working order, we find it to be
especially important to consider $I=3/2$ separately in order to access
$\bar{h}_1$. Because $\bar{h}_1$ starts to contribute from one order
higher in the $1/\nucleonmass$ series, we assume that our fits are
rather insensitive to this constant. The relation between the bare LEC
$h_1$ and the renormalized LEC $\bar{h}_1$ was given in
Sec.~\ref{sec:renormalization} (the same applies to $b_1$ and $\bar{b}_1$).
\subsection{Order \boldmath{${Q}^3$} results}	
In Table~\ref{tab:q3deltalessHB}, we show our fit results for the HB
LECs including uncertainties. For the $I=3/2$ ($I=1/2)$ fit, we used
$168$ $(112)$ data points, so that the reduced $\chi^2/n$ is equal to
$0.9$ ($1.8$), where $n$ stands for the number of data points minus
the number of fitted LECs. Table~\ref{tab:q3deltalesscov} collects the
corresponding results of the covariant approach, where we emphasize
again that the fits are fully comparable in terms of data
configuration and energy range. In the covariant formalism, we obtain
for the reduced $\chi^2/n$ $0.3$ ($0.2$) for the $I=3/2$ ($I=1/2)$
fit.

\begin{table}[htbp]
	\caption{Low-energy constants obtained from a $\deltapart$-less order-${Q}^3$ fit in the HB approach to the real parts of $s$- and $p$-wave photoproduction multipoles of the MAID model from Ref.~\cite{Drechsel:2007if}. All LECs are given in units of $\si{\per \square \giga\electronvolt}$.}
	\label{tab:q3deltalessHB}
	\centering
		\begin{tabular*}{\textwidth}{@{\extracolsep{\fill}}ccccc} 
		\hline
		&&&& \\[-7pt]
		& {$\hat{d}^r_8$} & {$\hat{d}^r_9$} &
		{$\hat{d}^r_{20}$} & {$\hat{d}^r_{21;22}$} \\[2pt]
		\hline
		&&&& \\[-7pt] 
		{\footnotesize HB order-${Q}^3$ fit value:} & \num{-7.7\pm 0.3} &	\num{0.03\pm 0.02}	&	\num{-17.2\pm 0.5}	&	\num{14.3\pm 0.5} \\[2pt]
		\hline
	\end{tabular*}
\end{table}

\begin{table}[htbp]
	\caption{Low-energy constants obtained from a $\deltapart$-less order-${Q}^3$ covariant fit to the real parts of $s$- and $p$-wave photoproduction multipoles of the MAID model from Ref.~\cite{Drechsel:2007if}. All LECs are given in units of $\si{\per\square \giga\electronvolt}$.}
	\label{tab:q3deltalesscov}
	\centering
		\begin{tabular*}{\textwidth}{@{\extracolsep{\fill}}ccccc} 
				\hline
				&&&& \\[-7pt]
				& {$\bar{d}_8$} & {$\bar{d}_9$} &
				{$\bar{d}_{20}$} & {$\bar{d}_{21;22}$} \\[2pt]
				\hline
				&&&& \\[-7pt] 
				{\footnotesize covariant order-${Q}^3$ fit value:} & \num{-4.9\pm 0.2} &	\num{ 0.01 \pm 0.01}	&	\num{-8.5\pm 0.3}	&	\num{9.4\pm 0.4} \\[2pt]
				\hline
		\end{tabular*}
\end{table}

\begin{figure}[tbp]
	\centering
	\includegraphics[width=\textwidth]{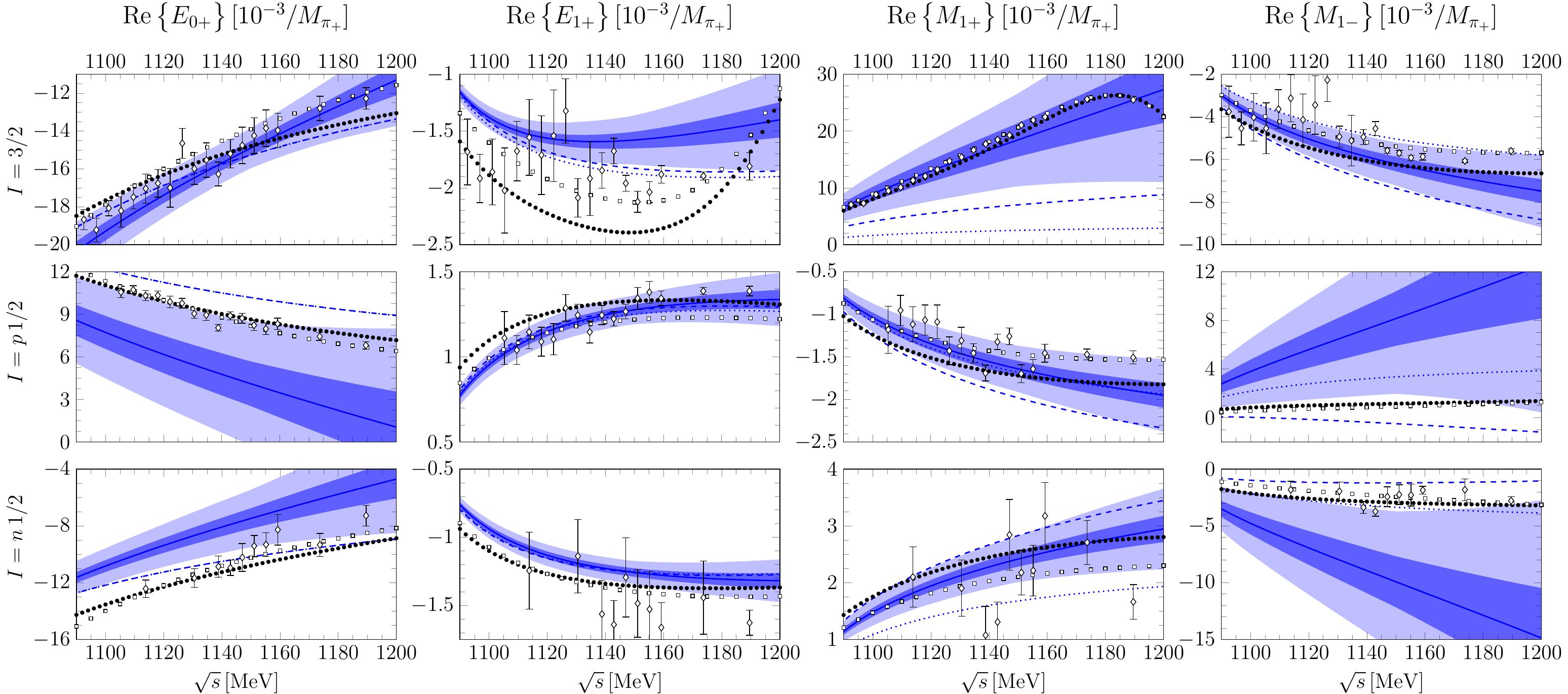}
	\caption{Order-${Q}^3$ fits obtained in the heavy baryon
          approach to the real parts of the $s$- and $p$-wave
          photoproduction multipoles. The solid, dashed and dotted
          lines denote the ${Q}^3$, ${Q}^2$ and ${Q}^1$ results,
          respectively. The darker (lighter) shaded bands show the
          estimated truncation errors at order ${Q}^3$ with $1\sigma$
          ($2\sigma$) confidence. The filled circles show the results
          of the MAID PWA from Ref.~\cite{Drechsel:2007if}, the
          squares (diamonds) are the results of the energy-dependent
          (independent) SAID analysis from
          Refs.~\cite{Briscoe:2019cyo, PCStrakovsky}
          (Ref.~\cite{Workman:2012jf}).}
	\label{fig:HBqTo3plot}
      \end{figure}

\begin{figure}[tbp]
	\centering
	\includegraphics[width=\textwidth]{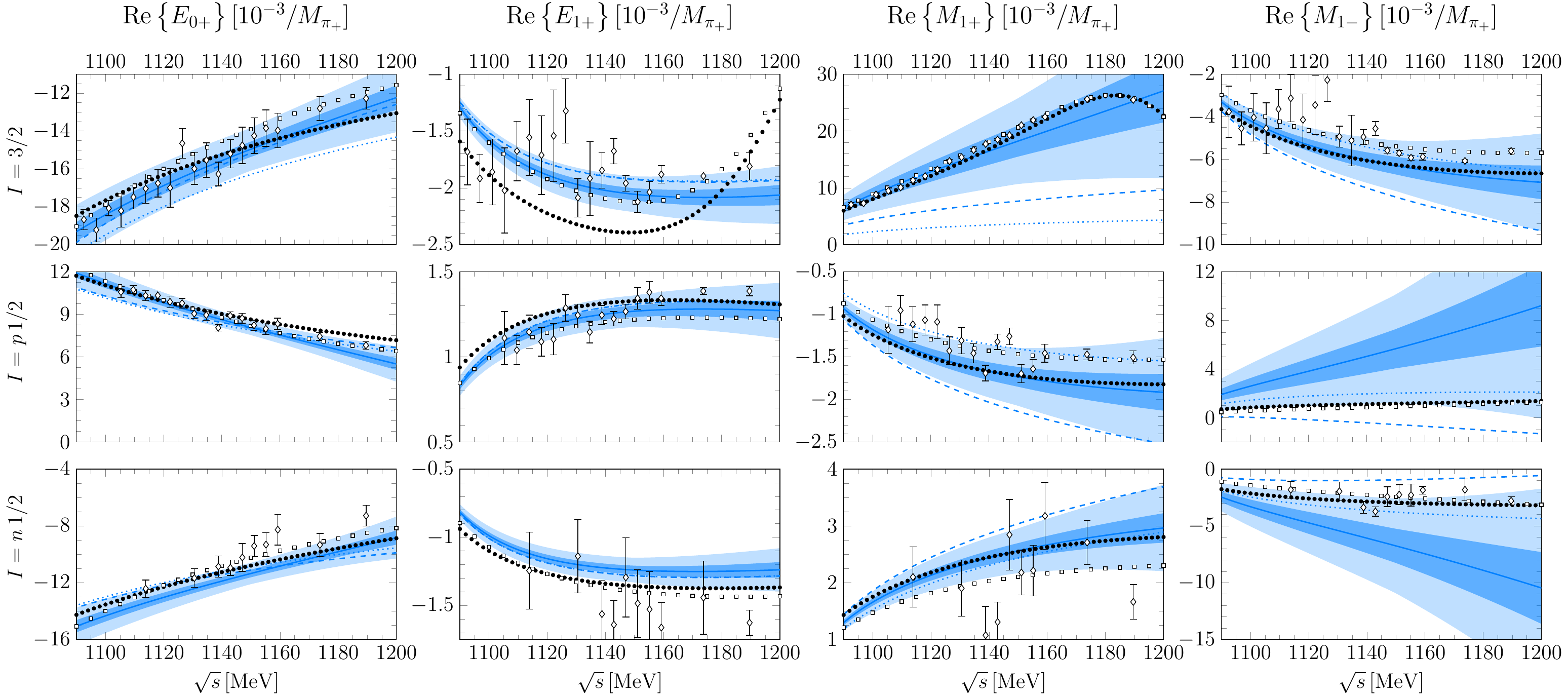}
	\caption{Order-${Q}^3$ fits obtained in the covariant approach to the real parts of the $s$- and $p$-wave photoproduction multipoles. The notation is as in Fig.~\ref{fig:HBqTo3plot}.}
	\label{fig:covqTo3plot}
\end{figure}

The HB (covariant) fit results are shown in Figs.~\ref{fig:HBqTo3plot}
(\ref{fig:covqTo3plot}), respectively, where the plotted energy range
corresponds to our fitting range. At this point, we remind that in the
$I=3/2$ channel, $E_{1+}^{3/2}$ has not been used in the fit. In both
$I=1/2$ channels, only $M_{1+}^{1/2}$ was used for the fit, all other
multipoles are predictions. We also remind that we used the
$1\sigma$ confidence interval for the determination of the truncation
errors, but our figures also show the $2\sigma$ band. Also, we adopted
a rather conservative value of the chiral breakdown scale 
$\Lambda_b=\SI{650}{\mega\electronvolt}$ motivated by recent studies
in the few-nucleon sector \cite{Epelbaum:2014efa, Epelbaum:2019zqc}.
Notice, however, that this estimation of $\Lambda_b$ in the
  few-nucleon sector
  \cite{Epelbaum:2014efa,Furnstahl:2015rha,Epelbaum:2019wvf} does not
  necessarily apply to the case at hand. In
  particular, for reactions where the $\deltapart$-resonance plays
  an important role such as e.g.~pion-nucleon scattering at
  intermediate energies, the breakdown scale of the $\deltapart$-less
  $\chi$PT is considerably lower, see e.g.~\cite{Siemens:2016hdi}.
      
In the $I=3/2$ channel, we find both HB and covariant descriptions
satisfactory with the exception of the electric $\deltapart$ multipole
$E_{1+}^{3/2}$, which is not well reproduced. The latter fact is
probably due to the missing $\deltapart$ contributions. We expect the
description of this amplitude to improve when including them, which
will be discussed in section~\ref{sec:qTo3epsTo2results}. Note that
the shape of the $M_{1+}^{3/2}$ multipole is not well reproduced for
CM energies above $\sqrt{s}=\SI{1150}{\mega\electronvolt}$, which is
also due to missing $\deltapart$ dynamics. We expect that the
inclusion of the  $\deltapart$ contributions will correct this
behaviour significantly (see also
Sec.~\ref{sec:qTo3epsTo2results}). The $E_{0+}^{3/2}$ and
$M_{1-}^{3/2}$ multipoles are better described in the covariant case,
which explains the differences in the fit quality. The smaller value
of $\chi^2/n$ in the covariant approach may indicate that the Bayesian
truncation uncertainties are overestimated or that the fit range is
not broad enough to constrain the LECs sufficiently. At this point we
emphasize that $n$ should not be equated with the number of degrees of
freedom, because our choice of the number of data points is in some
way arbitrary. Varying the energy steps results in different values of
the reduced $\chi^2/n$, therefore one cannot associate a perfect fit
with a value of $\chi^2/n=1$. The values of $\chi^2/n$ should,
therefore, not be directly interpreted as a statistical measure of the
fit quality but used to compare the relative
quality of fits obtained with the same preconditions.
Generally, one expects the
values of the multipoles at  neighboring energy points to be
correlated, so that the actual number of degrees of freedom is
presumably much smaller than $n$. For this reason, good fits of the
empirical data are expected to have  $\chi^2/n$ considerably smaller than $1$.


In the $I=1/2$ channels, the difference between HB and covariant approach is more pronounced. The descriptions of $E_{0+}^{1/2}$ is clearly better in the covariant approach, and also the description of $M_{1+}^{1/2}$ is better, which is reflected in the fit quality. Clearly, the value of $\chi^2/n=1.8$ in the HB case shows that the data cannot be well described. On the other hand, the small value of $\chi^2/n=0.2$ obtained in the covariant approach indicates that more data might be required to constrain $d_9$ better.

With the obtained fit values in both approaches, we now analyze the
numerical differences between HB and covariant constants as discussed
in Sec.~\ref{sec:LECs}. Substituting the employed values of $g_A$,
$F_\pi$  into the right-hand sides of Eq.~\eqref{eq:IRshifts}, the numerical values of the predicted shifts read
\begin{align}
	\notag & \Delta d_8^{\text{IR}} = \SI{0.6}{\per\square\giga\electronvolt},& \quad &\Delta d_9^{\text{IR}} = \SI{0.08}{\per\square\giga\electronvolt}, \\
	&\Delta d_{20}^{\text{IR}} = \SI{2.8}{\per \square\giga \electronvolt},& \quad &\Delta d_{21;22}^{\text{IR}} = \SI{-1.9}{\per\square\giga\electronvolt},
\end{align}
whereas we find for the actual differences from tables~\ref{tab:q3deltalessHB} and \ref{tab:q3deltalesscov}
\begin{align}
	\notag &\bar{d}_8-\hat{d}_8^r =\SI{2.8}{\per\square\giga\electronvolt},& \quad &\bar{d}_9-\hat{d}_9^r =\SI{-0.02}{\per\square\giga\electronvolt}, \\ 
	&\bar{d}_{20}-\hat{d}_{20}^r =\SI{8.7}{\per\square\giga\electronvolt},& \quad &\bar{d}_{21;22}-\hat{d}_{21;22}^r =\SI{-4.9}{\per\square\giga\electronvolt}.
\end{align}
These results show that the IR shifts can, at best, only
qualitatively explain the differences
between HB and covariant approach. In particular, the agreement for
$d_9$ is excellent, while for $d_8$, $d_{20}$ and $d_{21;22}$, the
differences have the same sign as the IR shift.
The remaining gap between the two sets of fit parameters is probably due to the poorer fit quality in the HB approach. We also find that the values of the covariant LECs are more natural as in the HB approach. Here, the term “natural” refers to the naive estimate that the $d_i$'s should roughly be of order one in the units of $\Lambda_b^{-2}$:
\begin{align}
	d_i \sim \frac{1}{\Lambda_b^2} \sim \SI{2.5}{\per \square \giga \electronvolt} \qq{with} \Lambda_b = \SI{650}{\mega \electronvolt}.
	\label{eq:naturalness}
\end{align}

\subsection{Order \boldmath{${Q}^3+\epsilon^2$} results}
\label{sec:qTo3epsTo2results}
In Table~\ref{tab:q3eps2covfit}, we show our fit results for the LECs in the covariant approach at order ${Q}^3+\epsilon^2$. The reduced $\chi^2/n$ is equal to $0.5$ $(1.0)$ in the $I=3/2$ ($I=1/2$) channel.
\begin{table}[htbp]
	\caption{Low-energy constants obtained from an order-${Q}^3+\epsilon^2$ covariant fit to the real parts of $s$- and $p$-wave photoproduction multipoles of the MAID model from Ref.~\cite{Drechsel:2007if}. All LECs are given in units of $\si{\per\square \giga\electronvolt}$, $b_1$ is given in $\nucleonmass^{-1}$.}
	\label{tab:q3eps2covfit}
	\centering
	\begin{tabular*}{\textwidth}{@{\extracolsep{\fill}}cccccc} 
		\hline
		&&&&& \\[-7pt]
		& {$\bar{d}_8$} & {$\bar{d}_9$} &
		{$\bar{d}_{20}$} & {$\bar{d}_{21;22}$} & {$b_1$} \\[2pt]
		\hline
		&&&&& \\[-7pt] 
		{\footnotesize covariant order-${Q}^3+\epsilon^2$ fit value:} & \num{0.72\pm 0.03} &	\num{0.02\pm 0.01}	&	\num{-2.1\pm 0.1}	&	\num{-0.1\pm 0.1} & \num{5.6\pm 0.1} \\[2pt]
		\hline
	\end{tabular*}
\end{table}

\begin{figure}[tbp]
	\centering
	\includegraphics[width=\textwidth]{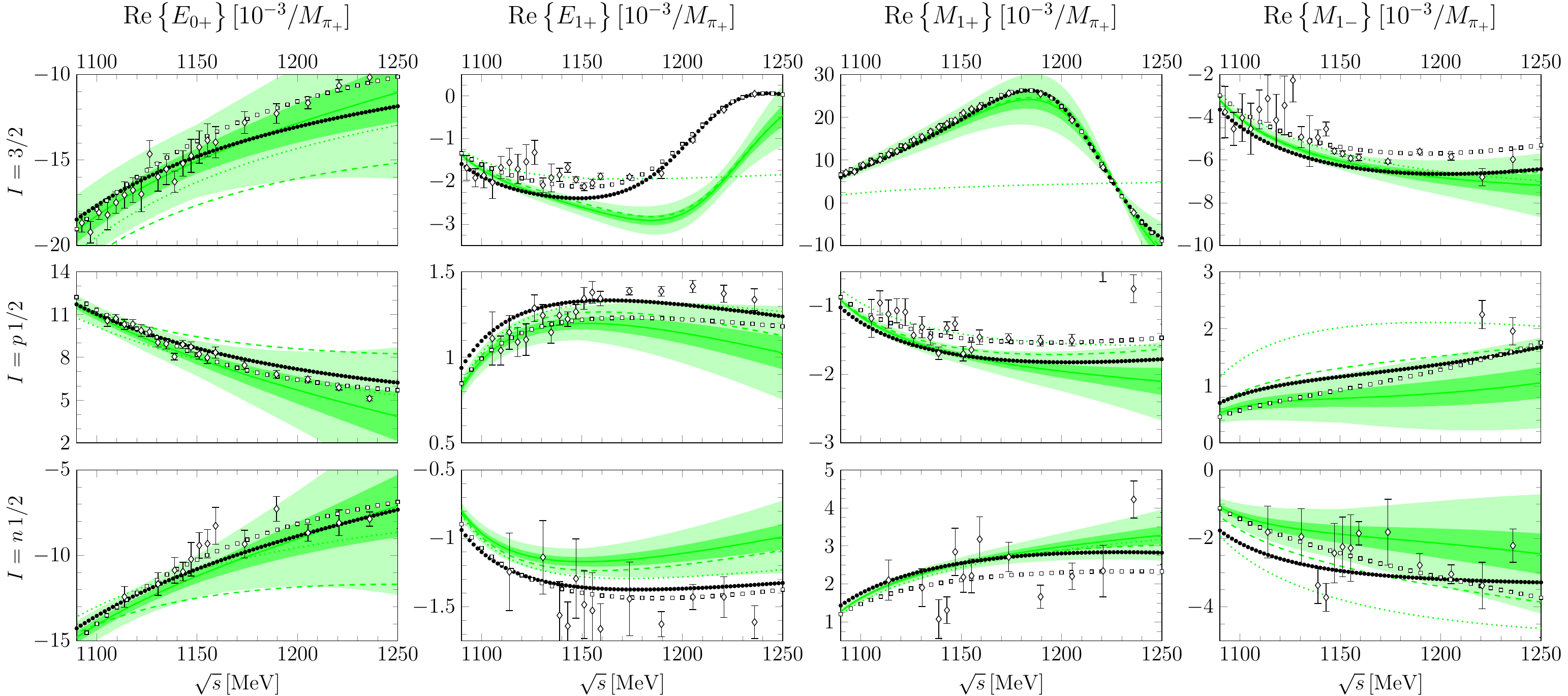}
	\caption{Order-${Q}^3+\epsilon^2$ fits obtained in the covariant approach to the real parts of the $s$- and $p$-wave photoproduction multipoles. The solid, dashed and dotted lines denote the ${Q}^3+\epsilon^2$, $\epsilon^2$ and $\epsilon^1$ results, respectively. The darker (lighter) shaded bands show the estimated truncation errors at order ${Q}^3$ with $1\sigma$ ($2\sigma$) confidence. The filled circles show the results of the MAID PWA from Ref.~\cite{Drechsel:2007if}, the squares (diamonds) are the results of the energy-dependent (independent) SAID analysis from Refs.~\cite{Briscoe:2019cyo, PCStrakovsky} (Ref.~\cite{Workman:2012jf}).}
	\label{fig:covqTo3epsTo2plot}
\end{figure}

In Fig.~\ref{fig:covqTo3epsTo2plot}, we show the results. The most
outstanding difference to the $\deltapart$-less case is the
significantly improved description of the $M_{1+}^{3/2}$ multipole. In
comparison to the $\deltapart$-less approach, the region beyond
$\sqrt{s}=\SI{1150}{\mega\electronvolt}$ is reproduced excellently
both in magnitude and in shape. The leading $\deltapart$ tree diagrams
suffice to correct the ${Q}^3$ description of $M_{1+}^{3/2}$. However,
the description of $E_{1+}^{3/2}$ is unsatisfactory. We presume that
including $\epsilon^3$ terms will improve the description, because the
subleading $\gamma N \Delta$ coupling constant $h_1$ contributes at
its leading order in $1/\nucleonmass$ to $E_{1+}^{3/2}$. In the
$I=1/2$ channels, for all $s$- and $p$-waves, the data description is not worse than
in the $\deltapart$-less case. Only $M_{1+}^{1/2}$ was fitted for comparability with the $\deltapart$-less approach, all other multipoles are reproduced to a good extent. Here, $M_{1-}^{1/2}$ must be accentuated, which was overshot dramatically in the $\deltapart$-less theory.

\begin{figure}[tbp]
	\centering
	\includegraphics[width=0.5 \textwidth]{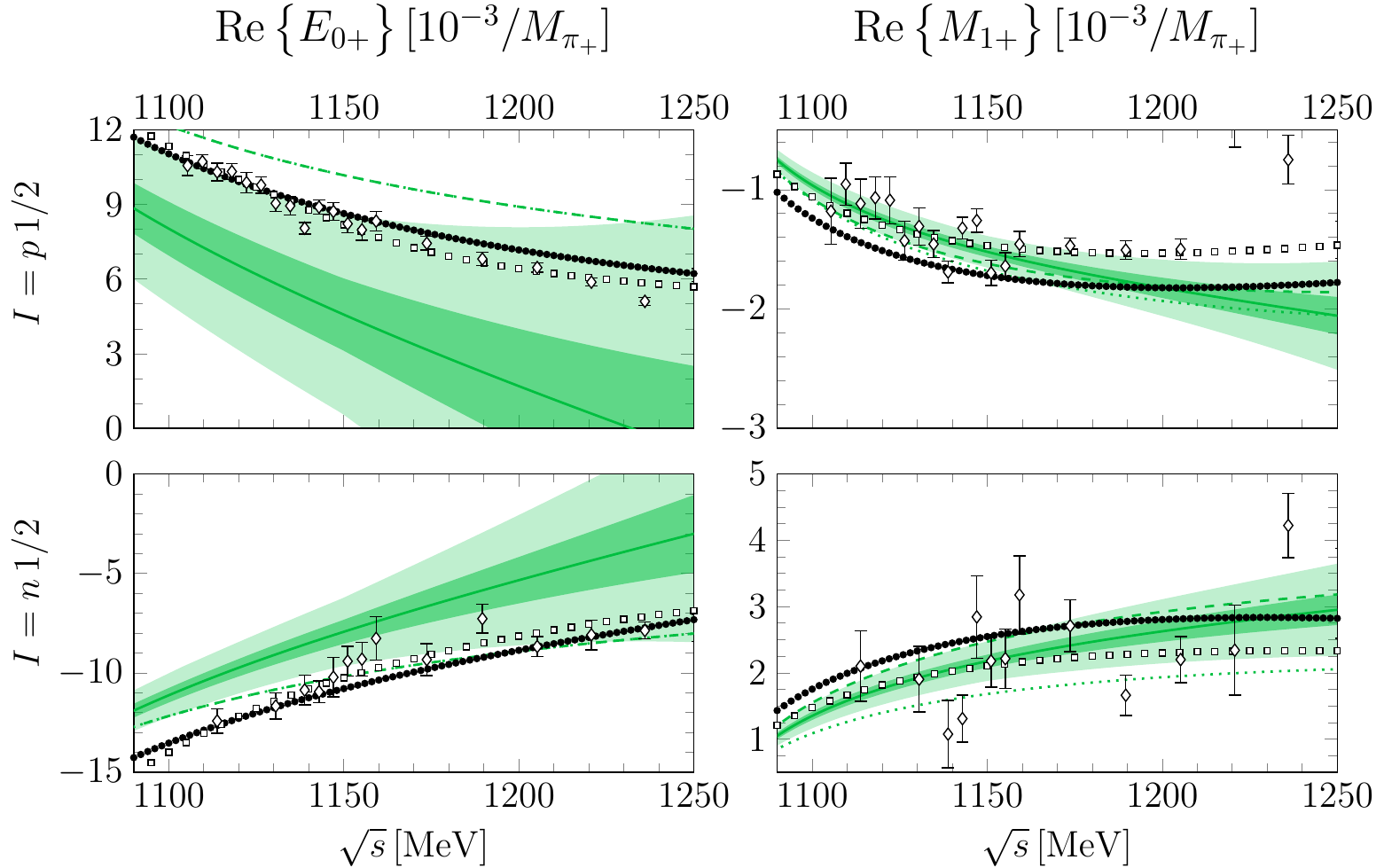}
	\caption{Selected plots of order-${Q}^3+\epsilon^2$ fits obtained in the heavy baryon approach to the real parts of the $s$- and $p$-wave photoproduction multipoles. The notation is as in Fig.~\ref{fig:covqTo3epsTo2plot}.}
	\label{fig:HBqTo3epsTo2plot}
\end{figure}

For the sake of completeness, we mention that we also performed fully
comparable fits in the HB approach, using a consistent value of the
$\deltapart$ mass fitted to the HB order-$\epsilon^2$ amplitude of
$\deltamass = (1196.1 -45.2\, \ci)\,\si{\mega \electronvolt}.$ In
Fig.~\ref{fig:HBqTo3epsTo2plot}, we show $E_{0+}$ and $M_{1+}$ in the
$I=1/2$ channel in order to illustrate that the HB results are,
however,  by far not as satisfying as the covariant results, which we observe for all multipoles.
We conclude that, as expected, the heavy-baryon (non-relativistic) expansion is not quite efficient 
if one wants to extend the scheme to the $\deltapart$ region.
Therefore, we do not give the full set of results here, however, we
found that the resulting value of $b_1= \num{5.5\pm 0.1} \nucleonmass^{-1}$ is still very close to the covariant one. This is a nice indication of the stability of this constant.

Next, we take a look at the differences between the $\deltapart$-less and -full fit values of the LECs from the point of $\deltapart$ resonance saturation, as explained in Sec.~\ref{sec:LECs}. Numerically, the expected differences read, using the covariant fit value of $b_1$
\begin{align}
	\delta d_8(\Delta)=-\delta d_{21;22}(\Delta) = \SI{-3.4}{\per\square\giga\electronvolt},
\end{align}
and for the actual differences
\begin{align}
	\notag	&\bar{d}_8^{\slashed{\Delta}}-\bar{d}_8^{\Delta} = \SI{-5.6}{\per\square\giga\electronvolt},& \quad &\bar{d}_9^{\slashed{\Delta}}-\bar{d}_9^{\Delta} = \SI{-0.01}{\per\square\giga\electronvolt},\\ &\bar{d}_{20}^{\slashed{\Delta}}-\bar{d}_{20}^{\Delta} = \SI{-6.4}{\per\square\giga\electronvolt},& \quad &\bar{d}_{21;22}^{\slashed{\Delta}}-\bar{d}_{21;22}^{\Delta} = \SI{9.5}{\per\square\giga\electronvolt}.
\end{align}
As one can see, the differences between $\deltapart$-less and
$\deltapart$-full parameters are only very qualitatively explained by the resonance saturation.

In the following, we compare our covariant order-${Q}^3+\epsilon^2$ results with data of the neutral pion production channel $\gamma p\to \pi^0  p$. Of the four physical reaction channels given in Eq.~\eqref{eq:physicalreactionchannels}, this channel is the most interesting for our purposes. The amplitudes for the two charged pion production channels $\gamma  p \to \pi^+ n$ and $\gamma n \to \pi^- p$
are dominated by the leading order Kroll-Ruderman terms, such that
subleading terms give only very small corrections. The remaining
neutral channel $\gamma n \to \pi^0 n$ is difficult to measure in
experiments since this requires a neutron target. Therefore, little
data are available for this channel. 

\begin{figure}[tbp]
	\centering
	\includegraphics[width=\textwidth]{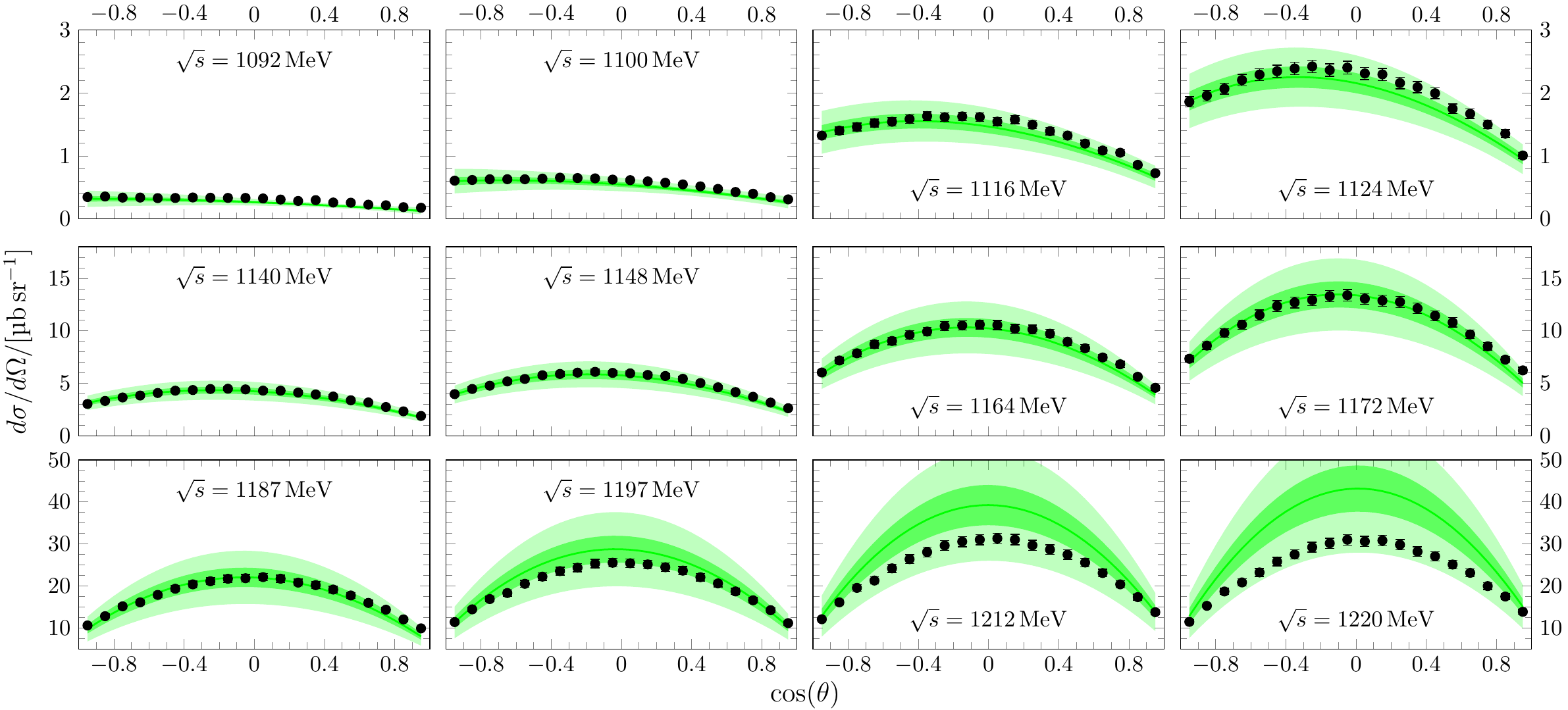}
	\caption{Covariant order-${Q}^3+\epsilon^2$ result of the unpolarized differential cross section in the channel $\gamma + p \to \pi^0 + p$. The solid lines denote the ${Q}^3+\epsilon^2$ results, the darker (lighter) shaded bands show the estimated truncation errors at order ${Q}^3+\epsilon^2$ with $1\sigma$ ($2\sigma$) confidence. The data are from Refs.~\cite{Hornidge:2012ca, PCHornidge}, error bars correspond to the combined statistical and systematical error.}
	\label{fig:crosssections}
\end{figure}

\begin{figure}[tbp]
	\centering
	\includegraphics[width=\textwidth]{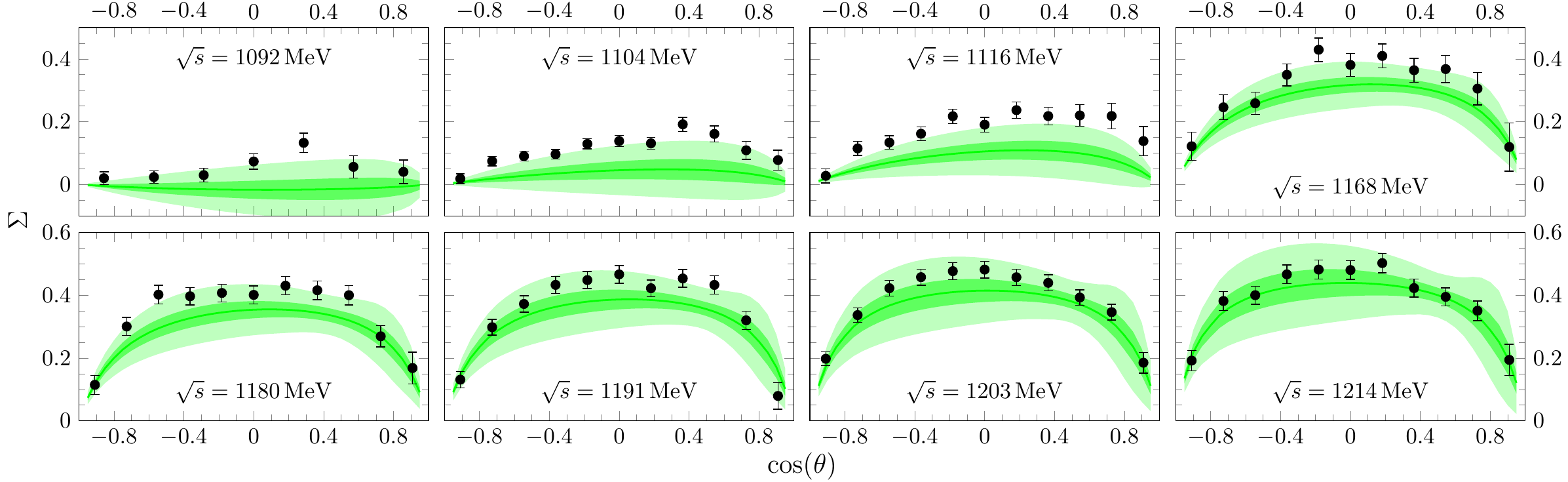}
	\caption{Covariant order-${Q}^3+\epsilon^2$ result of the linearly polarized photon asymmetry in the channel $\gamma + p \to \pi^0 + p$. The solid lines denote the ${Q}^3+\epsilon^2$ results, the darker (lighter) shaded bands show the estimated truncation errors at order ${Q}^3+\epsilon^2$ with $1\sigma$ ($2\sigma$) confidence. The data are from Refs.~\cite{Hornidge:2012ca, Hornidge:2013qka, PCHornidge}, error bars correspond to the combined statistical and systematical error.}
	\label{fig:asymmetries}
\end{figure}

The recent experiment at the Mainz Microtron (MAMI) provided high-precision data for the differential cross section $\dv{\sigma}{\Omega}$ and the linearly polarized photon asymmetry $\Sigma$ \cite{Hornidge:2012ca, Hornidge:2013qka, PCHornidge}. We compare our findings with these data and emphasize that our results for both observables were calculated as discussed in Sec.~\ref{sec:kinematics}, in particular we do not use the $s$- and $p$-wave approximation. In Figs.~\ref{fig:crosssections} and \ref{fig:asymmetries}, we depict our results obtained from the covariant order-${Q}^3+\epsilon^2$ fit, where the depicted error bars show the combined statistical and systematic error
\begin{align}
	\delta O_i =\sqrt{ \left( \delta O_i^{\text{stat}} \right)^2+\left( \delta O_i^{\text{sys}} \right)^2 }.
\end{align}
The systematic uncertainties are $\SI{4}{\percent}$ for $\dv{\sigma}{\Omega}$ and $\SI{5}{\percent}$ for $\Sigma$. 

In the analysis of the $s$- and $p$-wave multipole amplitudes, we found that the order-${Q}^3+\epsilon^2$ covariant fits give an overall accurate reproduction in our considered energy region. By comparison with the differential cross section and polarization asymmetry, we find our observation confirmed. Especially the polarization asymmetries are reproduced accurately up to the $\deltapart$ region. 

%
%
\subsection{Order \boldmath{$\epsilon^3$} results}
\begin{figure}[htbp]
	\centering
	\includegraphics[width=\textwidth]{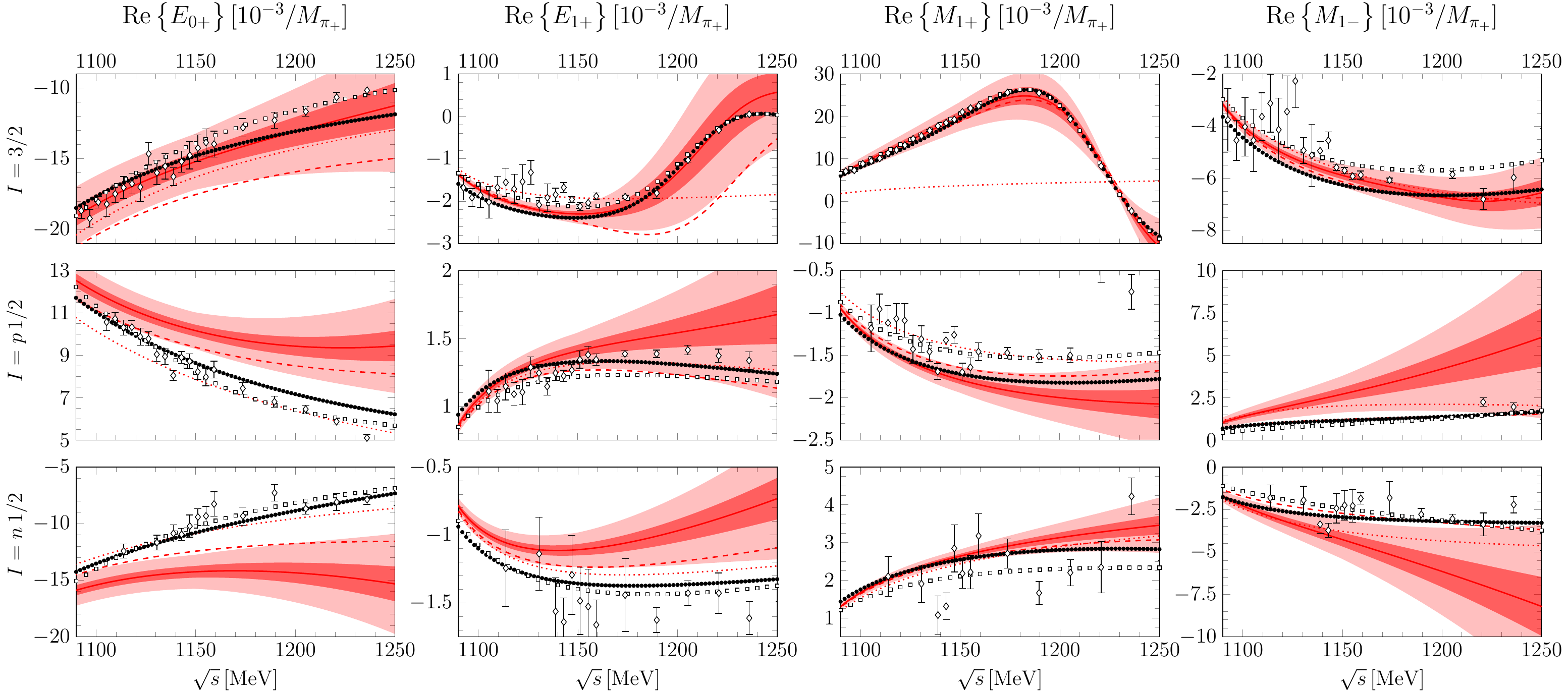}
	\caption{Order-$\epsilon^3$ fits obtained in the covariant approach to the real parts of the $s$- and $p$-wave photoproduction multipoles. The solid, dashed and dotted lines denote the $\epsilon^3, \epsilon^2$ and $\epsilon^1$ results, respectively. The darker (lighter) shaded bands show the estimated truncation errors at order ${Q}^3$ with $1\sigma$ ($2\sigma$) confidence. The filled circles show the results of the MAID PWA from Ref.~\cite{Drechsel:2007if}, the squares (diamonds) are the results of the energy-dependent (independent) SAID analysis from Refs.~\cite{Briscoe:2019cyo, PCStrakovsky} (Ref.~\cite{Workman:2012jf}).}
	\label{fig:covepsTo3plot}
\end{figure}

Finally, we give our results for the LECs determined from the covariant fit at order $\epsilon^3$ in Table~\ref{tab:eps3covfit}. In the $I=3/2$ fit, the number of used data points is $204$ due to the inclusion of the $E_{1+}$ multipole and $102$ in the $I=1/2$ channel. The reduced $\chi^2/n$ is equal to $0.2 (2.2)$ in the $I=3/2$($I=1/2$) channel. The corresponding results for the multipoles are depicted in Fig.~\ref{fig:covepsTo3plot}.
\begin{table}[htbp]
	\caption{Low-energy constants obtained from an order-$\epsilon^3$ covariant fit to the real parts of $s$- and $p$-wave photoproduction multipoles of the MAID model from Ref.~\cite{Drechsel:2007if}. All LECs are given in units of $\si{\per\square \giga\electronvolt}$, $\bar{b}_1$ and $\bar{h}_1$ are given in $\nucleonmass^{-1}$.}
	\label{tab:eps3covfit}
	\centering
	\begin{tabular*}{\textwidth}{@{\extracolsep{\fill}}ccccccc} 
		\hline
		&&&&&& \\[-7pt]
		& {$\bar{d}_8$} & {$\bar{d}_9$} &
		{$\bar{d}_{20}$} & {$\bar{d}_{21;22}$} & {$\bar{b}_1$} & {$\bar{h}_1$}\\[2pt]
		\hline
		&&&&&& \\[-7pt] 
		{\footnotesize covariant order-$\epsilon^3$ fit value:} & \num{-0.25 \pm 0.04} &	\num{-0.73 \pm 0.02}	&	\num{2.2 \pm 0.1} &	\num{-6.8 \pm 0.1} & \num{5.3\pm 0.1} & \num{0.9 \pm 0.1}\\[2pt]
		\hline
	\end{tabular*}
\end{table}

As we can see from Fig.~\ref{fig:covepsTo3plot}, the reproduction of the $I=3/2$ channel has improved compared to the covariant ${Q}^3+\epsilon^2$ fit. Especially $E_{1+}$ is now matched significantly better due to the inclusion of the subleading $\gamma N \Delta$ coupling constant $\bar{h}_1$. In the $I=1/2$ channels however, the description is distinctly worse compared to the ${Q}^3+\epsilon^2$-case. Especially the reproduction of the $E_{0+}^{1/2}$ multipoles fails, which has a substantial effect on the reproduction of cross sections, for example. Also, the fit quality in the $I=1/2$ channels of $\chi^2/n=2.2$ is significantly worse compared to the ${Q}^3+\epsilon^2$ fit, where we found $\chi^2/n=1.0$. We also remark that $M_{1-}^{1/2}$ is overshot again, but not as badly as in the $\deltapart$-less case (see Fig.~\ref{fig:covqTo3plot}). 
These observations indicate a slow convergence of the scheme in the $I=1/2$ channels, and one needs 
to extend the calculation to higher orders (such as ${Q}^4+\epsilon^3$ or $\epsilon^4$)
to obtain a better description of the data.

\begin{figure}[htbp]
	\centering
	\includegraphics[width=0.5 \textwidth]{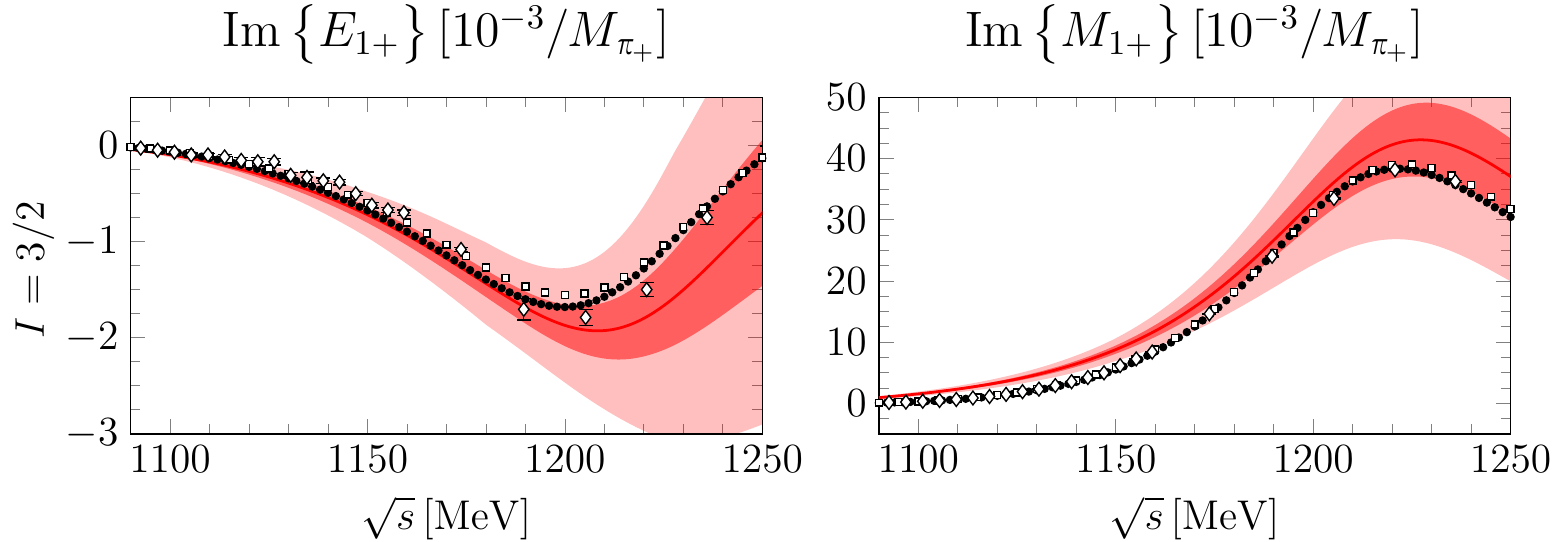}
	\caption{Imaginary parts of $E_{1+}^{3/2}$ and $M_{1+}^{3/2}$ using the covariant order-$\epsilon^3$ fit results. The notation is as in Fig.~\ref{fig:covepsTo3plot}.}
	\label{fig:covimplot}
\end{figure}

In order to check that our determined value of the $\deltapart$ mass in the covariant scheme is consistent with the $\deltapart$ contribution to the $\pi N$ elastic channel, we plot the imaginary part of $E_{1+}^{3/2}$ and $M_{1+}^{3/2}$ in Fig.~\ref{fig:covimplot}. Because the phase of the pion photoproduction amplitude is determined by the elastic $\pi N$ phase shifts, a satisfying reproduction of the imaginary parts is important. We find the agreement reasonable, with the deviation in $\Im \big\{M_{1+}^{3/2}\big\}$ close to threshold originating from the usage of the constant imaginary part of the $\deltapart$ mass in the $\deltapart$ pole diagrams. Furthermore, the $E2/M1$ ratio
\begin{align}
	R_{EM}=\frac{\Im \big\{E_{1+}^{3/2}\big\}}{\Im \big\{M_{1+}^{3/2}\big\} }\Bigg|_{\sqrt{s}=\SI{1232}{\mega\electronvolt}} \approx -0.033\pm 0.014
\end{align}
is consistent with the PDG value $-0.030 \lesssim R_{EM} \lesssim -0.020 $ \cite{Zyla:2020zbs}, whereas at order ${Q}^3+\epsilon^2$, we found $R_{EM} \approx -0.071\pm 0.014$.

%
%
%
%
\section{Summary and conclusions} 
\label{sec:summary}	
We have studied pion photoproduction in chiral effective field theory
with explicit $\deltapart$ degrees of freedom. Starting from the
$\deltapart$-less approach, we considered the reaction up to the
leading loop order in the heavy baryon and in the manifestly covariant
schemes. In particular, we analyzed the difference between the
obtained HB and covariant results for low-energy constants in terms of
the infrared regular shifts. We extended our calculations to the
leading $\deltapart$ contributions employing the complex-mass approach
using a fitted $\deltapart$ mass and studied the effects of resonance
saturation to the LECs. Moreover, we for the first time provide results for pion photoproduction at order $\epsilon^3$ in the small scale expansion scheme, where the leading $\deltapart$-full loop order is taken into account. The results for the LECs $\bar{d}_8, \bar{d}_9, \bar{d}_{20}, \bar{d}_{21;22}, \bar{b}_1$ and $\bar{h}_1$ are obtained by fits to the MAID partial-wave analysis using a Bayesian approach to theoretical uncertainties.

The main conclusions of our analysis of pion photoproduction can be summarized as follows:
\begin{itemize}
	\item In the $\deltapart$-less approach, the description of
          pion photoproduction is satisfying only in a very limited
          energy range above threshold and fails approaching the
          $\deltapart$ region. Especially for the magnetic multipole
          $M_{1+}^{3/2}$, the description agrees with the data only up
          to approximately $\sqrt{s}=\SI{1150}{\mega
            \electronvolt}$. Studying the reaction in the covariant
          framework yields a better agreement with the data than the
          heavy baryon approach. The results of our calculations
            are very relevant for ongoing investigations of
            few-nucleon electromagnetic reactions, see Ref.~\cite{Krebs:2020pii} for a
            review article. While the
            two-nucleon charge density operator at the leading
            one-loop order does not involve LECs from
            $\mathcal{L}_{\pi N}^{(3)}$
            \cite{Kolling:2009iq,Kolling:2011mt}, which allowed us to
            perform high-accuracy calculation of the deuteron charge
            and quadrupole form factors
            \cite{Filin:2019eoe,Filin:2020tcs}, the corresponding
            current operator depends on the LECs $d_8$, $d_9$,
            $d_{21}$ and $d_{22}$ \cite{Kolling:2011mt}. In
            particular, the LEC $d_9$ governs the long-range
            two-nucleon contribution to the deuteron magnetic form
            factor \cite{Kolling:2012cs}.
	\item Incorporating the leading $\deltapart$ tree
          contributions significantly extends the energy range in
          which a good agreement with the $s$- and $p$-wave multipoles
          can be achieved. We found that the leading $\gamma N \Delta$
          coupling constant $b_1$ is stable with respect to variation
          of the energy range, assigned relative error to the data and
          combination of $I=3/2$ and $I=1/2$ fit. The difference
          between the numerical values of the LECs obtained in a
          $\deltapart$-less and $\deltapart$-full approach can,
          at least very qualitatively,
          be explained in terms of resonance saturation.
          The results from the covariant order-${Q}^3+\epsilon^2$
          calculation are found to reproduce the high-precision data of cross sections and polarization asymmetries from Refs.~\cite{Hornidge:2012ca, Hornidge:2013qka, PCHornidge} remarkably well.
	On the other hand, the order-${Q}^3+\epsilon^2$  calculation performed within the heavy baryon scheme
	demonstrates a much worse description of the data. This is an indication of the fact that the $1/m_N$ expansion 
	is not efficient in the $\deltapart$ region.
	\item The next-to-leading $\deltapart$ contributions give rise
          to surprisingly large corrections to the scattering
          amplitude. However, these corrections are important to
          achieve a reasonable description of $E_{1+}^{3/2}$. At the
          same time, the description of the $I=1/2$ channel gets
          worsened significantly. The overall reproduction of the $s$-
          and $p$-wave multipoles is worse than in the
          ${Q}^3+\epsilon^2$ approach. However, the given estimate of
          the leading and subleading $\gamma N \Delta$ coupling
          constants $b_1$ and $h_1$ can be taken as reliable, because
          the isospin-$3/2$ channel is very well described. Also, the
          values agree with our findings from
          Ref.~\cite{Rijneveen:2020qbc}.
          Notice further that while the explicit treatment of the
  $\deltapart$-resonance in $\chi$PT helps to avoid the unnecessary
  lowering of the breakdown scale $\Lambda_b$, the expansion
  parameter in the $\deltapart$-scheme becomes $\propto \Delta $. It
  is, therefore, not a priori clear that the framework with explicit
  $\deltapart$ degrees of freedom features a smaller expansion
  parameter. For example, the $1/m_N$-expansion of the nucleon
  polarizabilities was found to converge considerably slower
  upon the explicit inclusion of the $\deltapart$-resonance
  \cite{Thurmann:2020mog}. Thus, the most efficient scheme can only be
  determined upon performing explicit calculations.  
\end{itemize}

Based on the conclusions of our analysis of pion photoproduction, we find that it would be very interesting to extend the analysis in the following points. In our work, we have focused on calculating the $s$- and $p$-wave multipoles, because they give by far the largest contributions to cross sections. However, in Refs.~\cite{FernandezRamirez:2009jb, FernandezRamirez:2009su}, the importance of $d$-waves to observables was pointed out. Therefore, it would be worthwhile to extend the analysis to higher partial waves or to the analysis of observables directly. Also, further insight could be gained from  
extending the covariant analysis to higher orders (${Q}^4+\epsilon^3$ or $\epsilon^4$)
given the fairly slow convergence of the small scale expansion
scheme. 
A $\deltapart$-less ${Q}^4$ calculation was already provided by Hilt et al.~\cite{Hilt:2013fda}, but the improvement in the description was only moderate, especially in the $\deltapart$ region. Since our analysis revealed significant improvement in the description of $I=3/2$ multipole amplitudes in the $\deltapart$ region, but a worse description of the $I=1/2$ channels, the effects of the ${Q}^4$ ($\epsilon^4$) terms in combination with the order-$\epsilon^3$ terms would be most interesting to study.

%
\section*{Acknowledgments} 
We are grateful to Igor Strakovsky for providing us the recent SAID solution for pion photoproduction and to David Hornidge for providing the full set of cross section and polarization asymmetry data from Refs.~\cite{Hornidge:2012ca,Hornidge:2013qka}.
This work was supported in part by BMBF (Grant No.~05P18PCFP1), by DFG and NSFC through funds provided to the
Sino-German CRC 110 ``Symmetries and the Emergence of Structure in QCD" (NSFC
Grant No.~12070131001, Project-ID 196253076 - TRR 110) and by DFG (Grant
No.~426661267).

	\newpage
	\appendix
\section{Relating different sets of amplitudes}
\label{sec:relationsofAisandBis} 	
The relation between the coefficients $A_i$ and $B_i$ of Eq.~\eqref{eq:ballamplitudes} and Eq.~\eqref{eq:minimalbasis} can be found by equating the two representations:
\begin{align}
	\sum_{i=1}^{8} B_i V_i^{\mu}=\sum_{i=1}^{4} A_i \mathcal{M}_i^{\mu}.
\end{align}
Remembering that $V_4$ and $V_8$ are not needed for real photons ($\epsilon\cdot k=0, k^2=0$) and using the two relations obtained by current conservation of the matrix element \eqref{eq:ballampsrelation}, the coefficients $A_i$ can be obtained from $B_i$ as follows:
\begin{align}
	A_1=\ci(B_5+\nucleonmass B_6), \quad A_2=\ci \frac{B_3}{k\cdot p+k\cdot p^\prime},\quad A_3=\ci B_7,\quad A_4=\frac{\ci}{2} B_6.
\end{align}
The relations between the invariant amplitudes $A_i$ and the amplitudes $\mathcal{F}_i$ read:
\begin{align}
	\notag \mathcal{F}_1 &= - \frac{W-\nucleonmass}{8\uppi W} \sqrt{(E_p+\nucleonmass)(E_{p^\prime}+\nucleonmass)} \left[ A_1 + (W-\nucleonmass) A_4 - \frac{2 \nu}{W-\nucleonmass} (A_3-A_4) \right], \\
	\notag \mathcal{F}_2 &= - \frac{W+\nucleonmass}{8\uppi W}\, \abs{\vec{q}}\, \sqrt{\frac{E_p - \nucleonmass}{E_{p^\prime}+\nucleonmass}} \left[ -A_1 + (W+\nucleonmass) A_4 - \frac{2 \nu}{W+\nucleonmass} (A_3-A_4) \right], \\ 
	\notag \mathcal{F}_3 &= - \frac{W+\nucleonmass}{8 \uppi W}\, \abs{\vec{q}}\, \sqrt{(E_p-\nucleonmass)(E_{p^\prime}+\nucleonmass)} \left[ \frac{W^2-\nucleonmass^2}{W+\nucleonmass} A_2 + A_3 - A_4 \right], \\
	\mathcal{F}_4 &= -\frac{W-\nucleonmass}{8\uppi W}\, \abs{\vec{q}}^2\, \sqrt{\frac{E_p + \nucleonmass}{E_{p^\prime}+\nucleonmass}} \left[ - \frac{W^2-\nucleonmass^2}{W-\nucleonmass} A_2 +A_3 - A_4 \right],
\end{align}
where we have used $\nu=-\tfrac{1}{2} k \cdot q$ and $W=\sqrt{s}$ is the CM energy.
%
%
%
%
\section{Feynman diagrams} \label{sec:Feynmandiagrams}
\begin{figure}[tbp]
	\centering
	\includegraphics[scale=.9]{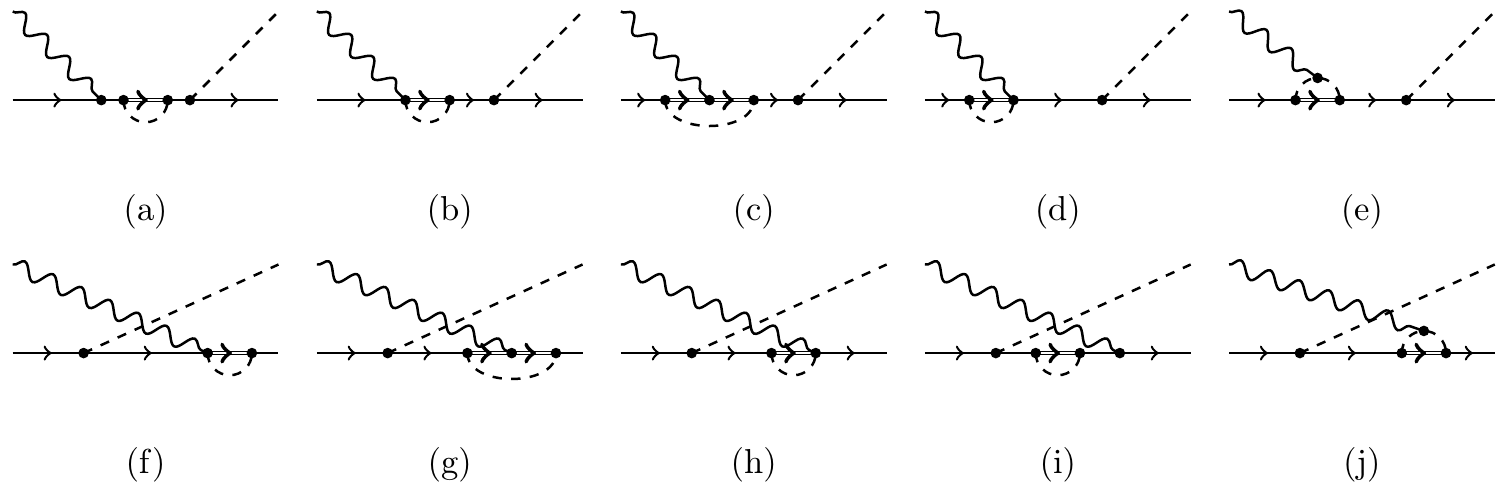}
	\caption{First set of $\deltapart$-full loop diagrams.\label{FG1}}
\end{figure}

\begin{figure}[htbp]
	\centering
	\includegraphics[scale=.9]{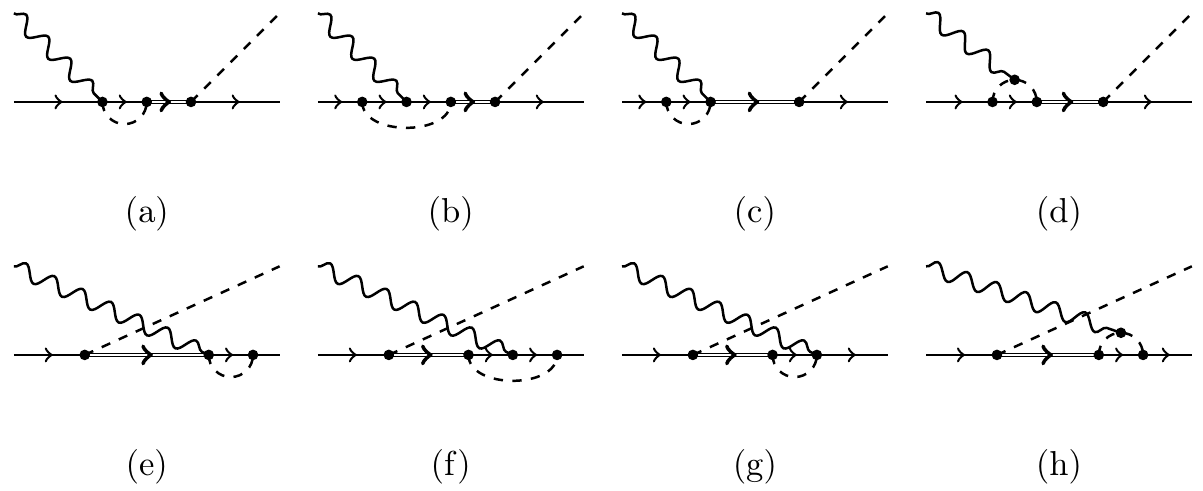}
	\caption{Second set of $\deltapart$-full loop diagrams. \label{FG2}}
\end{figure}

\begin{figure}[htbp]
	\centering
	\includegraphics[scale=.9]{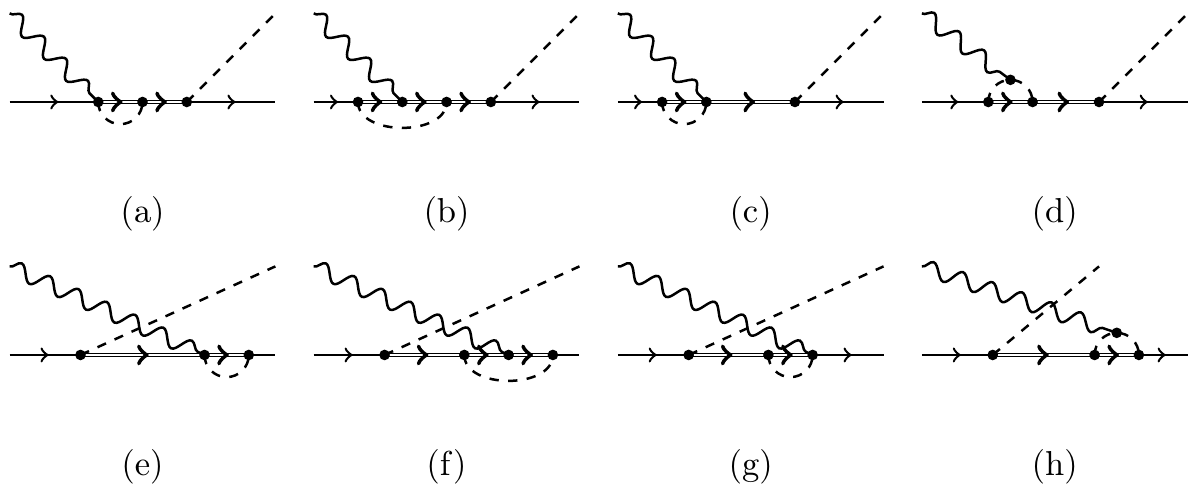}
	\caption{Third set of $\deltapart$-full loop diagrams. \label{FG3}}
\end{figure}

\begin{figure}[htbp]
	\centering
	\includegraphics[scale=.9]{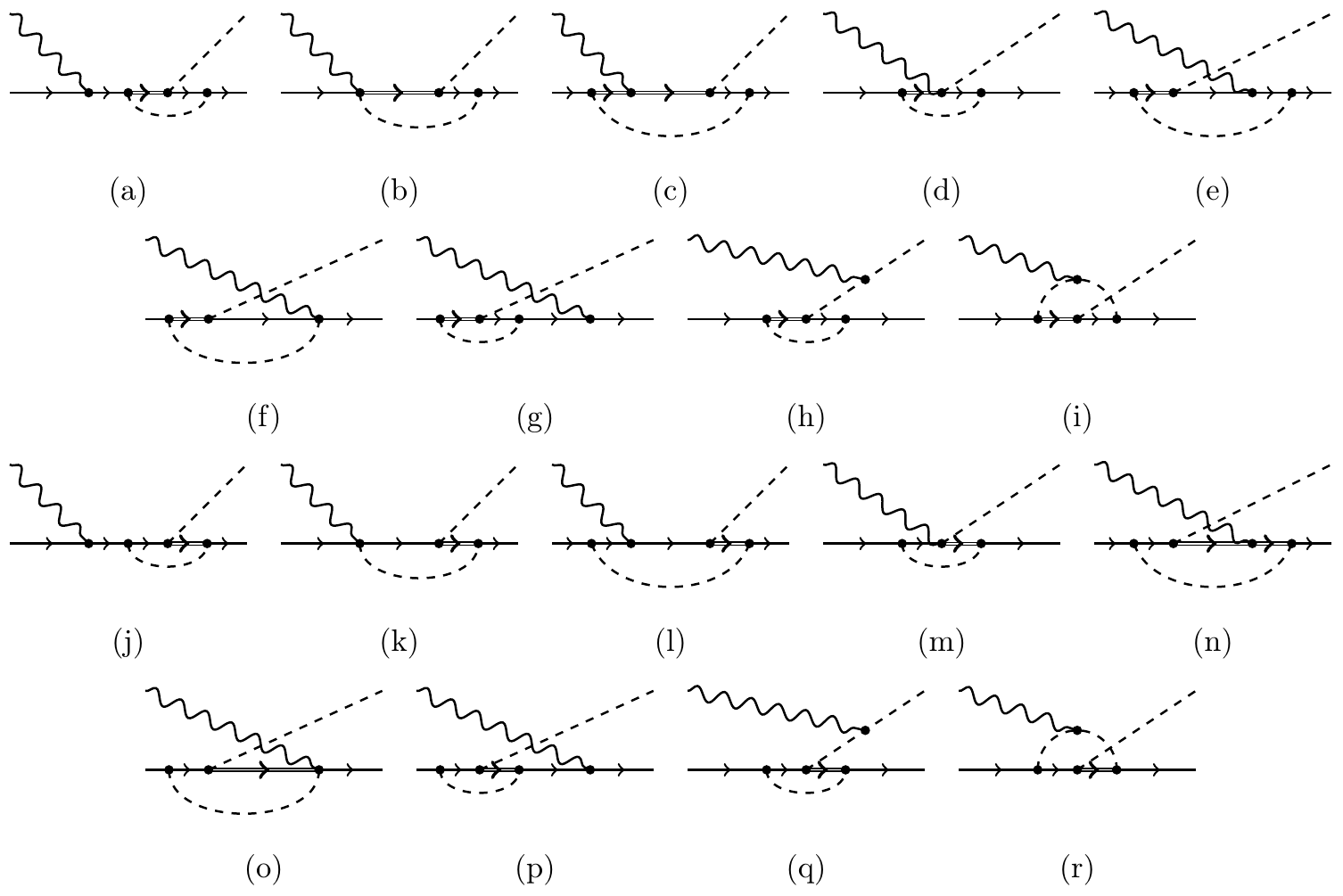}
	\caption{Fourth set of $\deltapart$-full loop diagrams. \label{FG4}}
	\label{fig:deltaloops}
\end{figure}

\begin{figure}[htbp]
	\centering
	\includegraphics[scale=.9]{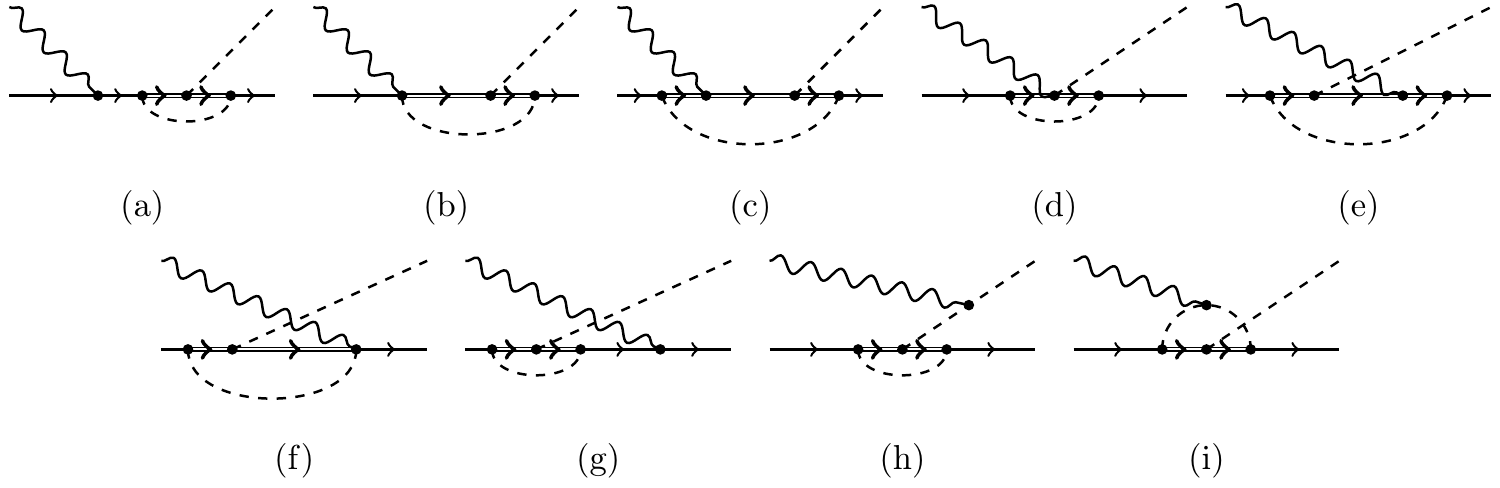}
	\caption{Fifth set of $\deltapart$-full loop diagrams. \label{FG5}}
\end{figure}

In Figs.~\ref{FG1},  \ref{FG2}, \ref{FG3}, \ref{FG4} and \ref{FG5}, we present the Feynman diagrams for pion photoproduction, which appear at order $\epsilon^3$. We cluster them in five gauge-invariant sets. The lower-order diagrams were already shown in Ref.~\cite{Rijneveen:2020qbc}.
%
%
%
%
\section{Counter terms} 
\label{sec:counterterms}
In this Appendix, we present the expressions for the renormalized quantities and the counter terms. 
\subsection{Loop integrals}\label{sec:loop_integrals}
The loop integral functions are defined as
\begin{align}
	\notag  A_0(m^2) & = \frac{1}{\ci} \int \frac{\dd^d l}{(2 \uppi)^d} \frac{\mu^{4-d}}{l^2-m^2},\\ 
	\notag B_0(p^2,m_0^2,m_1^2) & = \frac{1}{\ci} \int \frac{\dd^d l}{(2\uppi)^d} \frac{\mu^{4-d}}{(l^2-m_0^2) ((l+p)^2-m_1^2)},\\
	\notag J_0(\omega) & = \frac{1}{\ci} \int \frac{\dd^d l}{(2\uppi)^d} \frac{\mu^{4-d}}{(l^2-M_\pi^2)(v\cdot l + \omega)},\\
	C_0(p^2,(p-q)^2,q^2,m_0^2,m_1^2,m_2^2) & = \frac{1}{\ci} \int \frac{\dd^d l}{(2\uppi)^d} \frac{\mu^{4-d}}{(l^2-m_0^2)((l+p)^2-m_1^2)((l+q)^2-m_2^2)}.
\end{align}
\subsection{Mesonic counter terms}
The renormalization rules for the pion mass, field redefinition and decay constant are given below. We remind that the parameter $\upalpha$ is the unphysical constant from the general pion field parametrization (Eq.~\eqref{eq:generalpionfield}).
\begin{align}
	M^2 = &\, M_\pi^2+\delta M^{(4)}, &
	\delta M^{(4)} = & \, \frac{M_\pi^2}{2 \piondecayconstant^2} (A_0(M_\pi^2)-4M_\pi^2\, l_3), \\
	Z_\pi  = & \, 1+\delta Z_\pi^{(4)},& 
	\delta Z_\pi^{(4)} = &\,  \frac{1}{\piondecayconstant^2} (A_0(M_\pi^2) (1-10\upalpha) -2 M_\pi^2\, l_4) , \\
	F = &\,  \piondecayconstant+\delta \piondecayconstant^{(4)}, &
	\delta \piondecayconstant^{(4)} = & -\frac{1}{\piondecayconstant} (A_0(M_\pi^2)+M_\pi^2\,l_4). 
\end{align}
\subsection{Heavy baryon counter terms}
\paragraph{Nucleon mass and field renormalization}
The HB expressions for the redefinition of the nucleon mass and field read
\begin{align}
	m & =\nucleonmass+\delta m^{(2)}+\delta m^{(3)}, &
	\delta m^{(2)} & = 4 c_1 M_\pi^2, 
	& \delta m^{(3)} & = -\frac{3 \axialcoupling^2 \pionmass^2}{4 \piondecayconstant^2} J_0(0), \\
	Z_N & = 1+ \delta Z_N^{(3)},
	&	\delta Z_N^{(3)} & = -\frac{3 \axialcoupling^2 \pionmass^2}{32\, \uppi^2 \piondecayconstant^2} + \frac{9 \axialcoupling^2}{4 \piondecayconstant^2} A_0(\pionmass^2).
\end{align}
\paragraph{$\deltapart$ mass and field renormalization.}
The $\deltapart$ mass and field are renormalized as 
\begin{align}
 \baredeltamass &=  \deltamass + \kronecker
                  \deltamass^{(2)}+\dots\,,\quad \quad 
\kronecker \deltamass^{(2)} = 4c_1^{\deltapart} \pionmass^2\,,
\end{align}
and
\begin{align}
 Z_{\deltapart}&=1+\dots\,,
\end{align}
where the ellipses refer to terms which are not relevant at the
considered order in the expansion. 
\paragraph{Axial nucleon coupling}
The renormalization rules for the axial nucleon coupling constant
$\axialcoupling$ in the HB sector are given below. Note that we
already taken into account  the Goldberger-Treiman discrepancy to fully remove the
redundant for pion photoproduction constant $d_{18}$ from the rules.
For a recent high-precision determination of the pion-nucleon coupling
constants and the   Goldberger-Treiman discrepancy from neutron-proton
and proton-proton scattering data see Ref.~\cite{Reinert:2020mcu}. 
Here, $\baregA$ is the bare, $\axialcoupling$ is the physical constant.
\begin{align}
	\baregA  = \axialcoupling + \delta g^{(3)}, \quad	\delta g^{(3)}  = \left( -4 d_{16} + 2 d_{18} + \frac{\axialcoupling^3}{16\,\uppi^2 \piondecayconstant^2} \right)\pionmass^2 - \frac{(\axialcoupling+2 \axialcoupling^3)}{\piondecayconstant^2} A_0(\pionmass^2).
\end{align}
\paragraph{Electromagnetic form factors of the nucleon}
The replacement rules for the counter terms of the constants $c_6$ and $c_7$ are given below, where we denote the renormalized quantities by the bar.
\begin{align}
	c_6 & = \bar{c}_6 + \delta c_6^{(3)}, 
	&	\delta c_6^{(3)} & = -\frac{2 \nucleonmass \axialcoupling^2}{\piondecayconstant^2} J_0(0),\\
	c_7 & = \bar{c}_7 + \delta c_7^{(3)}, &
	\delta c_7^{(3)} & = \frac{\nucleonmass \axialcoupling^2}{\piondecayconstant^2} J_0(0).
\end{align}
\subsection{Covariant counter terms}
\paragraph{Nucleon mass and field renormalization}
In the following, we introduce the dimensionless parameters $\alpha=\frac{\pionmass}{\nucleonmass}$ and $\beta=\frac{\deltamass}{\nucleonmass}$. The ratio of the masses $\alpha$ is not to be confused with the unphysical off-shell parameter $\upalpha$. The renormalization rules for the nucleon mass and field redefinition are given below. For convenience, we give the contributions arising from the $\deltapart$ resonance separately. This means that all corrections $\delta x^{(i, \Delta)}$ are set to zero in the $\deltapart$-less case. 
\begin{align}
	m&= \nucleonmass+\delta \nucleonmass^{(2)}+\delta \nucleonmass^{(3)} + \delta \nucleonmass^{(3,\Delta)}, \\
	\delta \nucleonmass^{(2)} &= 4 \pionmass^2\, c_1, \\
	\delta \nucleonmass^{(3)} &= - \frac{3 \axialcoupling^2 \nucleonmass}{2 \piondecayconstant^2} (A_0(\nucleonmass^2)+\pionmass^2 B_0(\nucleonmass^2,\pionmass^2,\nucleonmass^2)), \\
	\notag \delta \nucleonmass^{(3,\Delta)} &= -\frac{\pindcoupling^2 \nucleonmass^3}{576\, \uppi^2 \piondecayconstant^2 \beta^2} \left[ \alpha ^4 (16 \beta +13)-4 \alpha ^2 (3 \beta +2)+3 \beta ^4-12 \beta ^3-4 \beta ^2+4
	\beta +2 \right]\\ 
	\notag & + \frac{\nucleonmass \pindcoupling^2}{6 \piondecayconstant^2 \beta^2} \Big[ \left(\alpha ^4-\alpha ^2 \left(2 \beta ^2-6 \beta -5\right)+(\beta -1) (\beta +1)^3 \right) A_0(\pionmass^2) \\
	\notag & \qquad \qquad - \left(\alpha ^4-2 \alpha ^2 \left(\beta ^2+\beta +1\right)+\beta ^4+2 \beta ^3-\beta ^2+2 \beta	 +1 \right) A_0(\deltamass^2) \\
	 & \qquad \qquad- \left(\alpha ^2-(\beta -1)^2\right) \left(\alpha ^2-(\beta +1)^2\right)^2 \nucleonmass^2\,  B_0(\nucleonmass^2,\pionmass^2,\deltamass^2) \Big], \\
	Z_N  &= 1+\delta Z_N^{(3)}+\delta Z_N^{(3,\Delta)},\\
	\delta Z_N^{(3)} &= \frac{3 \axialcoupling^2}{4 \piondecayconstant^2 (\alpha^2-4)} \Big[ \frac{\pionmass^2}{4\uppi^2}+ (5\alpha^2-12) A_0(\pionmass^2) - 4\alpha^2 A_0(\nucleonmass^2) - 4 \pionmass^2 (\alpha^2-3) B_0(\nucleonmass^2,\pionmass^2,\nucleonmass^2) \Big], \\
	\notag \delta Z_N^{(3,\Delta)} &= \frac{\pindcoupling^2}{6 \piondecayconstant^2 \beta^2} \Big[ \left(3 \alpha ^4-\alpha ^2 \left(6 \beta ^2+4 \beta +9\right)+(\beta +1)^2 (3 \beta ^2-2\beta +5)\right) A_0(\pionmass^2)\\
	\notag & \quad \quad - \left( 3 \alpha ^4+\alpha ^2 (-6 \beta ^2-4 \beta +2)+3 \beta ^4+4 \beta ^3+\beta^2-8 \beta -5 \right) A_0(\deltamass^2)\\
	\notag & \quad \quad - \Big( 3 \alpha ^6-\alpha ^4 (9 \beta ^2+4 \beta +1)-\alpha ^2 (-9 \beta ^4-8	\beta ^3-2 \beta ^2+4 \beta +7)  \\
	& \quad\quad-(\beta +1)^3 (3 \beta ^3-5 \beta ^2+7 \beta
	-5 ) \Big)  \nucleonmass^2\, B_0(\nucleonmass^2,\pionmass^2,\deltamass^2)\Big].
\end{align} 
\paragraph{$\deltapart$ mass and field renormalization.}
The $\deltapart$ mass and field are renormalized as 
\begin{align}
 \baredeltamass &=  \deltamass + \kronecker
                  \deltamass^{(2)}+\dots\,,\quad \quad 
\kronecker \deltamass^{(2)} = 4c_1^{\deltapart} \pionmass^2\,,
\end{align}
and
\begin{align}
 Z_{\deltapart}&=1+\dots\,.
\end{align}
where the ellipses refer to terms which are not relevant at the
considered order in the expansion. 
\paragraph{Axial nucleon coupling}
The renormalization rules for the axial nucleon coupling constant $\axialcoupling$ are given below. Note that we use the auxiliary variables $a_i, b_i, c_i$ only in this particular context for reasons of clarity and comprehensibility.
\begin{align}
	\baregA &= \, \axialcoupling+\delta g^{(3)}+\delta g^{(3,\Delta)},\\
	\notag\delta g^{(3)} &=  -2 (2 d_{16}-d_{18}) \pionmass^2 - \frac{3 \axialcoupling^3 \pionmass^2}{16 \uppi^2 \piondecayconstant^2(\alpha^2-4)} - \frac{\axialcoupling}{\piondecayconstant^2(\alpha^2-4)} \Big[ (\alpha^2-4+2 \axialcoupling^2(2\alpha^2-5)) A_0(\pionmass^2) \\
	\notag&\qquad \quad + (8-(2+3\axialcoupling^2)\alpha^2) A_0(\nucleonmass^2) + \axialcoupling^2 \nucleonmass^2(\alpha^2-4) B_0(\pionmass^2,\nucleonmass^2,\nucleonmass^2)\\
	& \qquad \quad + (8-2\alpha^2-3\axialcoupling^2(\alpha^2-3)) \pionmass^2\, B_0(\nucleonmass^2,\pionmass^2,\nucleonmass^2) + \axialcoupling^2 \pionmass^2 \nucleonmass^2 (\alpha^2-4) C_0(\nucleonmass^2,\pionmass^2,\nucleonmass^2,\pionmass^2,\nucleonmass^2,\nucleonmass^2)\Big],\\
	\notag\delta g^{(3,\Delta)} &= a_0+ a_1 A_0(\pionmass^2)+a_2 A_0(\nucleonmass^2)+a_3 A_0(\deltamass^2)+b_1 B_0(\pionmass^2, \nucleonmass^2, \deltamass^2) +b_2 B_0(\pionmass^2, \deltamass^2, \deltamass^2)\\
	\notag& + b_3 B_0(\nucleonmass^2,\pionmass^2,\nucleonmass^2) + b_4 B_0(\nucleonmass^2,\pionmass^2,\deltamass^2)+ c_1 C_0(\nucleonmass^2,\pionmass^2,\nucleonmass^2,\pionmass^2,\nucleonmass^2,\deltamass^2)\\
	&+c_2 C_0(\nucleonmass^2,\pionmass^2,\nucleonmass^2,\pionmass^2,\deltamass^2,\deltamass^2),\\
	\notag a_0&= -\frac{\axialcoupling \pindcoupling^2 \nucleonmass^2}{5184\, \uppi^2 \piondecayconstant^2\beta^2} \Big[\alpha ^4 (24 \beta +325)-4 \alpha ^2 \left(3 \beta ^4-6 \beta ^3-8 \beta
	^2+140 \beta +45\right) +12 \alpha ^6  -69 \beta ^4-384 \beta ^3+100 \beta ^2\\
	\notag & \qquad \quad +248 \beta +158 \Big]\\
	\notag & - \frac{5 g_1 \pindcoupling^2 \nucleonmass^2}{31104\, \uppi^2 \piondecayconstant^2 \beta^4} \Big[ 3 \alpha ^{10}-\alpha ^8 \left(3 \beta ^2+5 \beta +9\right)-\alpha ^6 \left(3 \beta^4-10 \beta ^2-30 \beta +60\right)\\
	\notag & \qquad \quad+\alpha ^4 \left(3 \beta ^6+5 \beta ^5-3 \beta ^4-20
	\beta ^3-114 \beta ^2-248 \beta -34\right)\\
	\notag & \qquad\quad -2 \alpha ^2 \left(7 \beta ^6+59 \beta
	^5-129 \beta ^4-59 \beta ^3-76 \beta ^2-68 \beta -2\right)\\
	& \qquad\quad -2 \beta  \left(17 \beta
	^5-82 \beta ^4+18 \beta ^3+28 \beta ^2+42 \beta +20\right) \Big],\\
	\notag a_1&=\,-\frac{\axialcoupling \pindcoupling^2}{54 \piondecayconstant^2 \beta^2} \Big[ 4 \alpha ^6+\alpha ^4 \left(-8 \beta ^2-8 \beta +31\right)+\alpha ^2 \left(4 \beta ^4+8 \beta ^3-46 \beta ^2-69\right)+23 \beta ^4 +4 \beta ^3 +12 \beta ^2+56 \beta +25 \Big] \\
	\notag & + \frac{5 g_1 \pindcoupling^2}{972 \piondecayconstant^2 \beta^4} \Big[ 3 \alpha ^{10}-\alpha ^8 \left(9 \beta ^2+5 \beta +10\right) +\alpha ^6 \left(9 \beta^4+10 \beta ^3+23 \beta ^2+17 \beta +6\right)\\
	\notag & \qquad \quad -\alpha ^4 \left(3 \beta ^6+5 \beta ^5+27\beta ^4+36 \beta ^3+55 \beta ^2+33 \beta -35\right)\\
	& \qquad \quad +2 \alpha ^2 \left(7 \beta ^6+17
	\beta ^5+17 \beta ^4-14 \beta ^3+44 \beta ^2+71 \beta +21\right) -2 (\beta +1)^2
	\left(7 \beta ^4-4 \beta ^3-9 \beta ^2+10 \beta +2\right)\Big],\\
	a_2&=\frac{2 \axialcoupling \pindcoupling^2}{27 \piondecayconstant^2 \alpha^2 \beta^2} \Big[ 5 \alpha ^6+3 \alpha ^4 (2 \beta +9)+\alpha ^2 \left(-3 \beta ^3-6 \beta ^2+33 \beta
	+16\right) +6 (\beta -1) (\beta +1)^2 \Big], \\
	\notag a_3&=\frac{\axialcoupling \pindcoupling^2}{54 \piondecayconstant^2 \alpha^2 \beta^2} \Big[ 4 \alpha ^8+\alpha ^6 \left(-8 \beta ^2-8 \beta +15\right)+\alpha ^4 \left(4 \beta ^4+8
	\beta ^3-50 \beta ^2-8 \beta +30\right)\\
	\notag & \qquad \quad +\alpha ^2 \left(23 \beta ^4+16 \beta ^3+13
	\beta ^2-92 \beta -73\right)-24 (\beta -1) (\beta +1)^2 \Big]\\
	\notag & + \frac{5 g_1 \pindcoupling^2}{972 \piondecayconstant^2 \beta^4} \Big[ 3 \alpha ^{10}-\alpha ^8 \left(9 \beta ^2+5 \beta +7\right)+\alpha ^6 \left(9 \beta ^4+10\beta ^3+17 \beta ^2+12 \beta -1\right)\\
	\notag & \qquad \quad -\alpha ^4 \left(3 \beta ^6+5 \beta ^5+24 \beta^4+31 \beta ^3+42 \beta ^2-15 \beta -15\right)\\
	& \qquad \quad +2 \alpha ^2 \left(7 \beta ^6+17 \beta^5+10 \beta ^4-5 \beta ^3+18 \beta ^2-25 \beta -7\right)-14 \beta ^6-20 \beta ^5+34 \beta ^4-20 \beta ^3+22 \beta ^2+28 \beta +4 \Big],\\
	b_1&=\frac{2 \axialcoupling \pindcoupling^2 \nucleonmass^2}{9 \piondecayconstant^2 \alpha^2 \beta^2} \Big[ \alpha ^6+\alpha ^4 \beta  (7 \beta +3)+\alpha ^2 \left(-2 \beta ^4+\beta ^3+\beta ^2-5\beta -3\right) +2 (\beta -1)^2 (\beta +1)^3 \Big], \\
	\notag b_2&=-\frac{5 g_1 \pindcoupling^2 \nucleonmass^2}{486 \piondecayconstant^2 \beta^3} \Big[ 5 \alpha ^6+\alpha ^4 \left(8 \beta ^2-6 \beta -11\right) -2 \alpha ^2 \left(8 \beta ^4+18\beta^3+13 \beta ^2-6 \beta -3\right) \\
	& \qquad \quad +4 \beta ^2 \left(3 \beta ^4+6 \beta ^3-2 \beta^2+6 \beta +3\right) \Big], \\
	\notag b_3&=\frac{2 \axialcoupling \pindcoupling^2 \nucleonmass^2}{27 \piondecayconstant^2 \alpha^2 \beta^2} \Big[ 5 \alpha ^8+2 \alpha ^6 (3 \beta +7)+\alpha ^4 \left(-3 \beta ^3-12 \beta ^2+24 \beta +5\right)\\
	& \qquad \quad +12 \alpha ^2 (\beta -1) (\beta +1)^2-6 \left(\beta ^5-\beta ^3+\beta^2-1\right) \Big], \\
	\notag b_4&= \frac{\axialcoupling \pindcoupling^2 \nucleonmass^2}{54 \piondecayconstant^2\alpha^2 \beta^2} \Big[ 4 \alpha ^{10}+\alpha ^8 \left(-12 \beta ^2-8 \beta +11\right)+\alpha ^6 \left(12 \beta^4+16 \beta ^3-61 \beta ^2+3\right)\\
	\notag & \qquad \quad +\alpha ^4 \left(-4 \beta ^6-8 \beta ^5+73 \beta^4+48 \beta ^3+2 \beta ^2-96 \beta -79\right) -\alpha ^2 (\beta +1)^2 \left(23 \beta^4-30 \beta ^3+26 \beta ^2+66 \beta -85\right)\\
	\notag & \qquad \quad +24 \left(\beta ^5-\beta ^3+\beta^2-1\right) \Big]\\
	\notag & + \frac{5 g_1 \pindcoupling^2 \nucleonmass^2}{972 \piondecayconstant^2 \beta^4} \Big[ 3 \alpha ^{12}-\alpha ^{10} \left(12 \beta ^2+5 \beta +10\right) +\alpha ^8 \left(18 \beta^4+15 \beta ^3+30 \beta ^2+17 \beta +6\right)\\
	\notag & \qquad \quad -\alpha ^6 \left(12 \beta ^6+15 \beta^5+44 \beta ^4+48 \beta ^3+54 \beta ^2-7 \beta -16\right)\\
	\notag & \qquad \quad +\alpha ^4 \left(3 \beta ^8+5 \beta ^7+38 \beta ^6+65 \beta ^5+76 \beta ^4+51 \beta ^3+44 \beta ^2-73 \beta -29\right)\\
	\notag & \qquad \quad -2 \alpha ^2 \left(7 \beta ^8+17 \beta ^7+17 \beta ^6-3 \beta ^5-12 \beta ^4+23 \beta ^3-7 \beta ^2-41 \beta -9\right)\\
	& \qquad \quad +2 (\beta +1)^3 \left(7 \beta ^5+\beta
	^4-17 \beta ^3+19 \beta ^2-8 \beta -2\right)\Big],\\
	\notag c_1&= \frac{4 \axialcoupling \pindcoupling^2 \nucleonmass^4}{9\piondecayconstant^2 \alpha^2 \beta ^2} \Big[ \alpha ^8+2 \alpha ^6 \left(2 \beta ^2+\beta -1\right)-2 \alpha ^4 \beta  \left(\beta^3+\beta +2\right) +2 \alpha ^2 (\beta -1)^2 (\beta +1)^3 \\
	&\qquad \quad -(\beta -1)^2 (\beta +1)^3 \left(\beta ^2-\beta +1\right)\Big], \\
	\notag c_2&= - \frac{5 g_1 \pindcoupling^2 \nucleonmass^4}{81 \piondecayconstant^2 \beta^3} \Big[ \alpha ^8+\alpha ^6 \left(2 \beta ^2-2 \beta -3\right)+\alpha ^4 \left(-6 \beta ^4-6\beta ^3-7 \beta ^2+4 \beta +3\right)\\
	&\qquad \quad +\alpha ^2 (\beta +1)^2 \left(5 \beta ^4-4 \beta^3+8 \beta ^2-1\right)-2 (\beta -1)^2 \beta ^2 (\beta +1)^4 \Big].
\end{align}
\paragraph{Electromagnetic form factors of the nucleon}
The renormalization rules of the two relevant LECs $c_6$ and $c_7$ are given below. We remind the reader that $\bar{c}_6$ and $\bar{c}_7$ are the renormalized quantities. Note that we use the auxiliary variables $a_i, b_i, c_i$ only in this particular context for reasons of clarity and comprehensibility.
\begin{align}
	c_6&=\bar{c}_6+\delta c_6^{(3)}+\delta c_6^{(3,\Delta)},\\
	\notag \delta c_6^{(3)}&= \frac{\axialcoupling^2}{(\alpha^2-4) \piondecayconstant^2} \Big[ \frac{\nucleonmass^2(4-3 \alpha^2)}{16\, \uppi^2} + (20-6\alpha^2) A_0(\pionmass^2) - 2 (8 - 3\alpha^2) A_0(\nucleonmass^2) \\
	& \qquad \quad + 2 \nucleonmass^2 (8 - 13 \alpha^2 + 3 \alpha^4) B_0(\nucleonmass^2,\pionmass^2,\nucleonmass^2) \Big],\\
	\delta c_6^{(3,\Delta)}=& \,a_0+ a_1 A_0(\pionmass^2)+a_2 A_0(\deltamass^2)+b_1 B_0(\nucleonmass^2,\pionmass^2,\deltamass^2),\\
	\notag a_0&=-\frac{\pindcoupling^2 \nucleonmass^2}{1296\, \uppi^2 \piondecayconstant^2 \beta^4} \Big[ \alpha ^4 \left(-27 \beta ^2+160 \beta +145\right)+20 \alpha ^2 \left(6 \beta ^3-7 \beta ^2-9 \beta -4\right)+27 \beta ^6\\
	& \qquad \quad +92 \beta ^5-281 \beta ^4-44\beta ^3+75 \beta ^2+60 \beta +20 \Big],\\
	\notag a_1&=-\frac{2 \pindcoupling^2}{81 \piondecayconstant^2\,\beta^4} \Big[\alpha ^4 \left(27 \beta ^2+20 \beta +15\right)+\alpha ^2 \left(-54 \beta ^4-40 \beta ^3-7 \beta ^2+120 \beta +75\right)\\
	& \qquad \quad +27 \beta ^6+20 \beta ^5-35 \beta^4+6 \beta ^3+47 \beta ^2-20 \beta -15 \Big],\\
	\notag a_2&= \frac{2 \pindcoupling^2}{81 \piondecayconstant^2 \beta^4} \Big[ \alpha ^4 \left(27 \beta ^2+20 \beta +15\right)-2 \alpha ^2 \left(27 \beta ^4+20 \beta ^3-10 \beta ^2+20 \beta +15\right)+27 \beta ^6\\ 
	&\qquad \quad +20 \beta ^5-62 \beta^4-86 \beta ^3-62 \beta ^2+20 \beta +15 \Big], \\
	\notag b_1&= \frac{2 \pindcoupling^2 \nucleonmass^2}{81 \piondecayconstant^2 \beta^4} \Big[ \alpha ^4 \left(27 \beta ^2+20 \beta +15\right)-2 \alpha ^2 \left(27 \beta ^4-7 \beta ^3-30 \beta ^2+5 \beta +15\right)+27 \beta ^6\\
	&\qquad \quad -34 \beta ^5-21 \beta^4+98 \beta ^3-57 \beta ^2-10 \beta +15 \Big] (-1+\alpha^2-2\beta -\beta^2),\\[1cm]
	c_7&=\bar{c}_7+\delta c_7^{(3)}+\delta c_7^{(3,\Delta)},\\
	\delta c_7^{(3)}&= \frac{\axialcoupling^2}{(\alpha^2 -4) \piondecayconstant^2} \Big[-\frac{\nucleonmass^2}{2\,\uppi^2} -4 A_0(\pionmass^2) +8 A_0(\nucleonmass^2) -4  \nucleonmass^2 (2-\alpha^2) B_0(\nucleonmass^2,\pionmass^2,\nucleonmass^2) \Big],\\
	\delta c_7^{(3,\Delta)}=& \,a_0+ a_1 A_0(\pionmass^2)+a_2 A_0(\deltamass^2)+b_1 B_0(\nucleonmass^2,\pionmass^2,\deltamass^2),\\
	a_0&= -\frac{\pindcoupling^2 \nucleonmass^2}{1296\, \uppi^2 \piondecayconstant^2 \beta^4} \Big[\alpha ^4 (32 \beta +29)+2 \alpha ^2 \left(12 \beta ^3-23 \beta ^2-18 \beta -8\right)+4 \beta ^5-67 \beta ^4+2 \beta ^3 +24 \beta ^2+12 \beta +4 \Big],\\
	a_1&= \frac{2 \pindcoupling^2}{81 \piondecayconstant^2 \beta^4} \Big[ \alpha ^4 (4 \beta +3)+\alpha ^2 \left(-8 \beta ^3+4 \beta ^2+24 \beta +15\right)+4 \beta ^5-7 \beta ^4-15 \beta ^3+4 \beta ^2 -4 \beta -3 \Big],\\
	a_2&= -\frac{2 \pindcoupling^2}{81 \piondecayconstant^2 \beta^4} \Big[ \alpha ^4 (4 \beta +3)-2 \alpha ^2 \left(4 \beta ^3-2 \beta ^2+4 \beta +3\right)+4 \beta ^5-7 \beta ^4-\beta ^3-7 \beta ^2 +4 \beta +3 \Big],\\
	\notag b_1&= \frac{2 \pindcoupling^2 \nucleonmass^2}{81 \piondecayconstant^2 \beta^4} (1-\alpha^2 + 2\beta +\beta^2) \Big[ \alpha ^4 (4 \beta +3)-2 \alpha ^2 \left(4 \beta ^3-6 \beta ^2+\beta +3\right)+4 \beta ^5-15 \beta ^4+25 \beta ^3-6 \beta ^2\\
	& \qquad \quad -2 \beta +3 \Big] .
\end{align}
\paragraph{$\pion \nucleon \deltapart$ coupling}
Here, we give the necessary shift of the $\pion \nucleon \deltapart$ coupling $\pindcoupling$:
\begin{align}
	\barehA=\pindcoupling+\delta \pindcoupling^{(2)}, \quad 
	\delta \pindcoupling^{(2)}=b_3(\nucleonmass-\deltamass)+b_6 \frac{\pionmass^2+\nucleonmass^2-\deltamass^2}{2\nucleonmass}.
\end{align}
\paragraph{Electromagnetic transition form factors}
Here, we give the shift for the coupling constants $b_1$ and $h_1$. Note that the auxiliary variables $a_i, b_i$ and $c_i$ are used only in this particular context. Taking the real part of the corrections to $b_1$ and $h_1$ ensures that the bare LECs are real, which is necessary for the Lagrangian to be hermitian. 
\begin{align}
	b_1&= \bar{b}_1+\delta b_1^{(3)},\\
	\notag \delta b_1^{(3)}&= \operatorname{Re}\Big[a_0+a_1 A_0(\pionmass^2)+a_2 A_0(\nucleonmass^2)+a_3 A_0(\deltamass^2) + b_1 B_0(\nucleonmass^2,\pionmass^2,\nucleonmass^2) + b_2 B_0(\nucleonmass^2,\pionmass^2,\deltamass^2) \\
	\notag &+ b_3 B_0(\deltamass^2,\pionmass^2,\nucleonmass^2)+ b_4 B_0(\deltamass^2,\pionmass^2,\deltamass^2) + c_1 C_0(0,\deltamass^2,\nucleonmass^2,\pionmass^2,\pionmass^2,\nucleonmass^2)+ c_2 C_0(0,\deltamass^2,\nucleonmass^2,\pionmass^2,\pionmass^2,\deltamass^2)\\
	&+c_3 C_0(\nucleonmass^2,0,\deltamass^2,\pionmass^2,\nucleonmass^2,\nucleonmass^2)+c_4 C_0(\nucleonmass^2,0,\deltamass^2,\pionmass^2,\deltamass^2,\deltamass^2)\Big],\\
	\notag a_0&= 2\nucleonmass(\beta-1) h_{15} + \nucleonmass(\beta^2-1) h_{16} - \frac{\axialcoupling \pindcoupling \nucleonmass}{16\,\uppi^2 \piondecayconstant^2(\beta^2-1)} \Big[ 2 \alpha ^2+\beta ^2-3 \beta \Big]\\
	\notag & + \frac{5 g_1 \pindcoupling \nucleonmass}{10368\,\uppi^2 \piondecayconstant^2 \beta^4 (\beta^2-1)} \Big[ \alpha ^6 \left(3 \beta ^3-2 \beta ^2-3 \beta +2\right) \beta -\alpha ^4 \left(3 \beta^6+3 \beta ^5+6 \beta ^4+117 \beta ^3+3 \beta ^2+8 \beta +4\right) \\
	\notag &\qquad \quad -\alpha ^2 \left(3	\beta ^8-2 \beta ^7+\beta ^6+74 \beta ^5-130 \beta ^4-264 \beta ^3+36 \beta^2-6\right)  +3 \beta ^{10} +3 \beta ^9 -6 \beta ^8-617 \beta ^7-153 \beta ^6\\
	&\qquad \quad +766 \beta^5 -84 \beta ^4-82 \beta ^3+26 \beta ^2+2 \beta -2 \Big],\\
	\notag a_1&= \frac{\axialcoupling \pindcoupling}{\piondecayconstant^2\, \nucleonmass (\beta-1)} + \frac{5 g_1 \pindcoupling}{324 \piondecayconstant^2\, \nucleonmass \beta^6 (\beta^2-1)} \Big[ \alpha ^6 \beta ^3 \left(3 \beta ^3-2 \beta ^2-3 \beta +2\right)\\
	\notag &\qquad \quad -\alpha ^4 \left(9 \beta^8-\beta ^7+\beta ^6-7 \beta ^5-2 \beta ^4-2 \beta ^2+2\right) +\alpha ^2 \beta ^2	\left(9 \beta ^8+4 \beta ^7+5 \beta ^5+43 \beta ^4+\beta ^3-8 \beta ^2+22 \beta-4\right)\\
	& \qquad \quad -\beta ^3 \left(3 \beta ^9+3 \beta ^8-4 \beta ^7-\beta ^6+32 \beta ^5+13	\beta ^4-68 \beta ^3  +75 \beta ^2+25 \beta -6\right) \Big],\\
	a_2&= \frac{\axialcoupling \pindcoupling (1-3\beta)}{\piondecayconstant^2\, \nucleonmass (\beta^2-1)},\\
	\notag a_3&=- \frac{5 g_1 \pindcoupling}{324 \piondecayconstant^2\, \nucleonmass\beta^6 (\beta^2-1)} \Big[ \alpha ^6 \beta ^3 \left(3 \beta ^3-2 \beta ^2-3 \beta +2\right) -\alpha ^4 \left(9 \beta^8-\beta ^7-2 \beta ^6-5 \beta ^5+\beta ^4-2 \beta ^3-2 \beta ^2+2\right)\\
	\notag &\qquad \quad +\alpha ^2	\beta ^3 \left(9 \beta ^7+4 \beta ^6-6 \beta ^5+4 \beta ^4+42 \beta ^3-82 \beta ^2+75	\beta +26\right)\\
	&\qquad \quad -\beta ^3 \left(3 \beta ^9+3 \beta ^8-7 \beta ^7-4 \beta ^6+36 \beta^5-62 \beta ^4-5 \beta ^3 +21 \beta ^2+81 \beta +6\right) \Big],\\
	b_1&= -\frac{\axialcoupling \pindcoupling \nucleonmass}{\piondecayconstant^2\, (\beta^2-1)^2} \Big[ \alpha ^2 \left(\beta ^3-2 \beta ^2-2 \beta -1\right)+2 \beta  \left(\beta ^2-\beta+2\right) \Big],\\
	\notag b_2&= - \frac{5 g_1 \pindcoupling \nucleonmass}{324 \piondecayconstant^2\,\beta^3 (\beta^2-1)^2} \Big[ \alpha ^8 (3 \beta -2) \left(\beta ^2-1\right)^2 -\alpha ^6 \left(12 \beta ^6-3 \beta^5-14 \beta ^4-2 \beta ^3+13 \beta -6\right) \beta \\
	\notag &\qquad \quad +\alpha ^4 \left(18 \beta ^8+3 \beta ^7-20 \beta ^6-3 \beta ^5+46 \beta ^4-47 \beta ^3-104 \beta ^2-41 \beta+4\right) \beta \\
	\notag &\qquad \quad +\alpha ^2 \left(-12 \beta ^{11}-7 \beta ^{10}+22 \beta ^9+4 \beta^8-84 \beta ^7-10 \beta ^6+152 \beta ^5+72 \beta ^4 +80 \beta ^3+65 \beta ^2-2 \beta+8\right)\\
	& \qquad \quad +3 \beta ^{13}+3 \beta ^{12}-10 \beta ^{11}-7 \beta ^{10}+43 \beta ^9+36	\beta ^8-172 \beta ^7-6 \beta ^6-23 \beta ^5 +19 \beta ^4-46 \beta ^3+33 \beta ^2-11\beta -6 \Big], \\
	b_3&=-\frac{\axialcoupling \pindcoupling \nucleonmass \beta}{\piondecayconstant^2\, (\beta^2-1)^2} \Big[ \alpha ^2 (3 \beta +1)+\beta ^3-5 \beta ^2+\beta -1 \Big],\\
	\notag b_4&=- \frac{5 g_1 \pindcoupling \nucleonmass}{162 \piondecayconstant^2\,\beta^6 (\beta^2-1)^2} \Big[ \alpha ^6 \left(4 \beta ^5-3 \beta ^4-2 \beta ^2+1\right)+2 \alpha ^4 \beta ^2 \left(11\beta ^5+11 \beta ^4+17 \beta ^3+4 \beta ^2-6 \beta -1\right)\\
	&\qquad \quad +2 \alpha ^2 \beta ^4 \left(23 \beta ^5-23 \beta ^4-62 \beta ^3-21 \beta ^2+6 \beta +5\right) +36 \left(3\beta ^{10}-\beta ^9-\beta ^8+\beta ^6\right) \Big],\\
	c_1&=-\frac{2 \axialcoupling \pindcoupling \nucleonmass^3 \alpha^2(\alpha^2+\beta^2-1)}{\piondecayconstant^2\, (\beta^2-1)},\\
	c_2&=- \frac{10 g_1 \pindcoupling \nucleonmass^3}{9 \piondecayconstant^2\, \beta (\beta^2-1)} \Big[ \alpha ^6-\alpha ^4 \left(\beta ^2-\beta +1\right)+3 \alpha ^2 \beta ^2 \left(\beta^2-1\right) \Big],\\
	c_3&=-\frac{2 \axialcoupling \pindcoupling \nucleonmass^3 \beta}{\piondecayconstant^2\,(\beta^2-1)} \Big[ \alpha ^2-\beta ^2+\beta -2 \Big],\\
	c_4&= \frac{10 g_1 \pindcoupling \nucleonmass^3}{27 \piondecayconstant^2\,(\beta^2-1)} \Big[ \alpha ^4 (\beta -4)-\alpha ^2 \left(4 \beta ^3-8 \beta ^2-\beta -4\right) -3 \left(3\beta ^4-\beta ^3-\beta ^2+1\right) \Big],\\[1cm]
	h_1&=\bar{h}_1+\delta h^{(3)},\\
	\notag \delta h^{(3)}&= \operatorname{Re}\Big[a_0+a_1 A_0(\pionmass^2)+a_2 A_0(\nucleonmass^2)+a_3 A_0(\deltamass^2) + b_1 B_0(\nucleonmass^2,\pionmass^2,\nucleonmass^2) + b_2 B_0(\nucleonmass^2,\pionmass^2,\deltamass^2) \\
	\notag & + b_3 B_0(\deltamass^2,\pionmass^2,\nucleonmass^2)+ b_4 B_0(\deltamass^2,\pionmass^2,\deltamass^2) + c_1 C_0(0,\deltamass^2,\nucleonmass^2,\pionmass^2,\pionmass^2,\nucleonmass^2)+ c_2 C_0(0,\deltamass^2,\nucleonmass^2,\pionmass^2,\pionmass^2,\deltamass^2)\\
	&+c_3 C_0(\nucleonmass^2,0,\deltamass^2,\pionmass^2,\nucleonmass^2,\nucleonmass^2)+c_4 C_0(\nucleonmass^2,0,\deltamass^2,\pionmass^2,\deltamass^2,\deltamass^2)\Big],\\
	\notag a_0&=2 \nucleonmass h_{15} - \frac{\axialcoupling \pindcoupling \nucleonmass(\alpha^2-\beta)}{4\,\uppi^2 \piondecayconstant^2\, (\beta-1)^2(\beta+1)}\\
	\notag &+ \frac{5 g_1 \pindcoupling \nucleonmass}{10368\,\uppi^2\, \piondecayconstant^2 \beta^4 (\beta-1)^2(\beta+1)} \Big[ 3 \alpha ^6 (\beta -1)^2 (\beta +1) \beta  -\alpha ^4 \left(3 \beta ^6+2 \beta ^5+7 \beta^4+275 \beta ^3+2 \beta ^2-5 \beta +4\right)\\
	\notag &\qquad \quad -\alpha ^2 \left(3 \beta ^8-3 \beta^7+\beta ^6-45 \beta ^5-222 \beta ^4-376 \beta ^3+56 \beta ^2+16 \beta -6\right)\\
	&\qquad \quad +3\beta ^{10}+2 \beta ^9-5 \beta ^8-587 \beta ^7+74 \beta ^6+717 \beta ^5 -386 \beta^4-138 \beta ^3+28 \beta ^2+6 \beta -2 \Big],\\
	\notag a_1&=\frac{2 \axialcoupling \pindcoupling}{\piondecayconstant^2\, \nucleonmass (\beta-1)^2}  + \frac{5 g_1 \pindcoupling}{324 \piondecayconstant^2\, \nucleonmass\, \beta^6 (\beta-1)^2(\beta+1)} \Big[ 3 \alpha ^6 (\beta -1)^2 \beta ^3 (\beta +1)\\
	\notag &\qquad \quad -\alpha ^4 \left(9 \beta ^8-4 \beta ^7+2\beta ^6-12 \beta ^5-3 \beta ^4+10 \beta ^3-2 \beta ^2-2 \beta +2\right)\\
	\notag & \qquad \quad +\alpha ^2\beta ^2 \left(9 \beta ^8+\beta ^7+2 \beta ^6+\beta ^5+48 \beta ^4+36 \beta ^3+45\beta ^2+30 \beta -28\right)\\
	& \qquad \quad -\beta ^3 \left(3 \beta ^9+2 \beta ^8-3 \beta ^7+\beta^6+35 \beta ^5+83 \beta ^4-51 \beta ^3 +37 \beta ^2+40 \beta -3\right) \Big],\\
	a_2&= - \frac{4 \axialcoupling \pindcoupling \beta}{\piondecayconstant^2\, \nucleonmass(\beta-1^2)(\beta+1)},\\
	\notag a_3&= - \frac{5 g_1 \pindcoupling}{324 \piondecayconstant^2\, \nucleonmass\, \beta^6 (\beta-1)^2(\beta+1)} \Big[ 3 \alpha ^6 (\beta -1)^2 \beta ^3 (\beta +1) -\alpha ^4 \left(9 \beta ^8-4 \beta ^7-\beta^6-9 \beta ^5+7 \beta ^3-2 \beta ^2-2 \beta +2\right)\\
	\notag &\qquad \quad +\alpha ^2 \beta ^2 \left(9 \beta^8+\beta ^7-4 \beta ^6+2 \beta ^5+46 \beta ^4-44 \beta ^3+129 \beta ^2+29 \beta-24\right)\\
	&\qquad \quad -\beta ^3 \left(3 \beta ^9+2 \beta ^8-6 \beta ^7-\beta ^6+38 \beta ^5+132	\beta ^4+\beta ^3  -136 \beta ^2+108 \beta +3\right) \Big],\\
	b_1&= -\frac{2 \axialcoupling \pindcoupling \nucleonmass}{\piondecayconstant^2\, (\beta-1)^3 (\beta+1)^2} \Big[ \alpha ^2 \left(\beta ^3-2 \beta ^2-2 \beta -1\right)+\beta  \left(\beta ^2+3\right) \Big],\\
	\notag b_2&= - \frac{5 g_1 \pindcoupling \nucleonmass}{324 \piondecayconstant^2\,\beta^3 (\beta-1)^3 (\beta+1)^2} \Big[ 3 \alpha ^8 (\beta -1)^3 (\beta +1)^2 -\alpha ^6 \left(12 \beta ^6-7 \beta ^5-13 \beta^4-2 \beta ^3-2 \beta ^2+17 \beta -5\right) \beta \\
	\notag & \qquad \quad +\alpha ^4 \left(18 \beta ^9-3 \beta^8-17 \beta ^7-5 \beta ^6+47 \beta ^5-115 \beta ^4-143 \beta ^3 -79 \beta ^2+3 \beta +6\right)\\
	\notag & \qquad \quad -\alpha ^2 \left(12 \beta ^{10}+3 \beta ^9-19 \beta ^8+88 \beta ^6-70 \beta^5-262 \beta ^4-120 \beta ^3  -80 \beta ^2-125 \beta -3\right) \beta \\
	& \qquad \quad +3 \beta ^{13}+2	\beta ^{12}-9 \beta ^{11}-3 \beta ^{10}+44 \beta ^9+5 \beta ^8-262 \beta ^7-98 \beta^6 +15 \beta ^5+84 \beta ^4-65 \beta ^3+13 \beta ^2-14 \beta -3 \Big], \\
	b_3&= - \frac{2 \axialcoupling \pindcoupling \nucleonmass \beta}{\piondecayconstant^2\, (\beta-1)^3(\beta+1)^2} \Big[ \alpha ^2 (3 \beta +1)-3 \beta ^2-1 \Big],\\
	\notag b_4&= \frac{5 g_1 \pindcoupling \nucleonmass}{162 \piondecayconstant^2\, \beta^6 (\beta-1)^3 (\beta+1)^2} \Big[ \alpha ^6 \left(\beta ^5+3 \beta ^4-6 \beta ^3+2 \beta ^2+\beta -1\right)\\
	\notag & \qquad \quad -2 \alpha ^4	\beta ^2 \left(43 \beta ^5+33 \beta ^4+13 \beta ^3-15 \beta ^2-7 \beta +5\right)\\
	& \qquad \quad +2\alpha ^2 \beta ^4 \left(23 \beta ^5+69 \beta ^4+43 \beta ^3-4 \beta ^2+12 \beta+1\right) -12 \beta ^6 \left(2 \beta ^5+13 \beta ^4-4 \beta ^2-2 \beta +3\right) \Big],\\
	c_1&=- \frac{2 \axialcoupling \pindcoupling \nucleonmass^3 \alpha^3(2 \alpha^2+\beta^2-1)}{\piondecayconstant^2 (\beta-1)^2(\beta+1)},\\
	c_2&=- \frac{10 g_1 \pindcoupling \nucleonmass^3 \alpha^2}{9 \piondecayconstant^2\,\beta (\beta-1)^2(\beta+1)} \Big[ 2 \alpha ^4+\alpha ^2 \left(-\beta ^2+\beta -2\right)+\beta  \left(2 \beta ^3-\beta ^2-2
	\beta +1\right) \Big],\\
	c_3&=- \frac{2 \axialcoupling \pindcoupling \nucleonmass^3}{\piondecayconstant^2 (\beta-1)^2(\beta+1)} \Big[ \alpha ^2 (\beta +1)-\beta  \left(\beta ^2+3\right) \Big],\\
	c_4&= -\frac{10 g_1 \pindcoupling \nucleonmass^3}{27 \piondecayconstant^2\,(\beta-1)^2(\beta+1)} \Big[2 \alpha ^4 (\beta +2)-\alpha ^2 \left(4 \beta ^3+11 \beta ^2-\beta +4\right) +2 \beta^5+13 \beta ^4-4 \beta ^2-2 \beta +3 \Big].
\end{align}

	\bibliographystyle{myunsrtnat}
	\bibliography{lit}
	
\end{document}